\documentclass[onecolumn,authoryear]{els-mrw} 

\usepackage{amsmath,amssymb,amsfonts,amsthm,makeidx,graphicx}
\usepackage{txfonts}
\usepackage{helvet}

\newcommand{\MLg}{\ensuremath{\Upsilon_\mathrm{gas}}}
\newcommand{\Vg}{\ensuremath{V_\mathrm{gas}}}
\newcommand{\MLd}{\ensuremath{\Upsilon_\mathrm{disk}}}
\newcommand{\Vd}{\ensuremath{V_{*\mathrm{disk}}}}
\newcommand{\MLb}{\ensuremath{\Upsilon_\mathrm{bulge}}}
\newcommand{\Vb}{\ensuremath{V_{*\mathrm{bulge}}}}

\begin{document}

\chapter{Modified Newtonian Dynamics (MOND)}\label{chap1}

\author[1]{Benoit Famaey}
\author[1,2]{Amel Durakovic}
\address[1]{\orgname{Universit\'e de Strasbourg, CNRS UMR 7550}, \orgdiv{Observatoire astronomique de Strasbourg}, \orgaddress{11 rue de l'Universit\'e, 67000 Strasbourg, France}}
\address[2]{\orgname{CEICO -- FZU}, \orgdiv{Institute of Physics of the Czech Academy of Sciences}, \orgaddress{Na Slovance 1999/2, 182 00 Prague 8, Czechia}}

\articletag{Chapter Article tagline}

\maketitle

\begin{glossary}[Glossary]
\term{External Field Effect} A change in the internal dynamics of a system when this system is immersed into an external gravitational field. \\
\term{Modified gravity} That gravitation is not according to general relativity, or its weak-field, slow-motion limit, Newtonian gravity. \\
\term{Modified inertia} That the force of inertia $m a$ is modified for low accelerations $a \ll a_0$, and is rather $m a^2/a_0$ (for circular motion at least). \\
\term{Phantom density} Fictitious density to
interpret MOND behaviour in Newtonian terms. Being fictitious, it
need not be always positive. \\
\term{Radial Acceleration Relation} An observed algebraic relation
between the dynamically inferred radial acceleration 
and the radial acceleration expected in Newtonian gravity from the
observed baryonic distribution. \\
\end{glossary}

\begin{glossary}[Nomenclature]
\begin{tabular}{@{}lp{34pc}@{}}
AeST & Aether-Scalar-Tensor \\
AQUAL & A-quadratic Lagrangian \\
BIMOND & Bi-metric MOND \\
BSTV & Bi-Scalar-Tensor-Vector \\
BTFR & Baryonic Tully-Fisher Relation\\
CMB & Cosmic Microwave Background \\
CSDR (or CDR) & Central (Surface) Densities Relation \\
EFE & External Field Effect \\
EMOND & Extended MOND \\
GQUMOND & Generalized Quasi-linear MOND \\
GR   & General relativity \\
HSB & High surface brightness \\
LSB & Low surface brightness \\
$\Lambda$CDM & $\Lambda$ Cold Dark matter \\
MOND & Modified Newtonian Dynamics (or Milgromian Dynamics)\\
PCG & Phase Coupling Gravitation \\
QUMOND & Quasi-linear MOND \\
RAR & Radial Acceleration Relation \\
TeVeS & Tensor-Vector-Scalar \\
TriMOND & Tripotential MOND \\
\end{tabular}
\end{glossary}

\begin{abstract}[Abstract]

This encyclopedia chapter presents Modified Newtonian Dynamics (MOND), the proposal that, below a certain acceleration scale $a_0$, dynamics departs from the Newtonian expectation, offering an alternative solution to the missing mass problem for galactic systems. In that context, the determining factor for the appearance of missing matter is predicted to be the acceleration, and not the mass or size of the system. MOND enables, for example, the prediction of rotation curves from only the baryonic distribution of galaxies. The simple rule is that the acceleration observed in the low-acceleration regime is the square root of the Newtonian expectation times $a_0$. Immediately, the flatness of rotation curves follows, as well as the proportionality of the fourth power of the asymptotic circular speed to only the baryonic mass of the galaxy. While the asymptotic circular speed is predicted not to depend on the baryonic surface density of galaxies of fixed baryonic mass, the inner shape of rotation curves is predicted to strongly depend on it. More generally, MOND implies an algebraic relation between the acceleration expected from Newtonian gravity and the total observed acceleration, at any radius in a galaxy. This is known, observationally, as the Radial Acceleration Relation. For galaxy clusters, it is commonly accepted that MOND fails, needing a stronger gravitational force (or more baryonic mass than observed) to account for the thermodynamic state of galaxy clusters, their lensing and kinematics. MOND, however, is not a complete theory, but a phenomenological non-relativistic paradigm in the limit of low accelerations, in need of embedding in a more fundamental theory. While various non-relativistic field theories of MOND exist, the search for a relativistic theory that recovers general relativity for high accelerations and MOND for low accelerations in the quasi-static limit, as well as a cosmology compatible with observations, is still on-going.
\end{abstract}

\newpage 

\begin{abstract}[Key points]
\begin{itemize}
    \item MOND is the proposal that there is a transition to new dynamics for accelerations below $a_0 \simeq 10^{-10} {\rm m}/{\rm s}^2$, acting as a new fundamental constant. In this low-acceleration regime, the equations of motion for point masses, or a continuous limit thereof, become spacetime scale-invariant.
    \item Applying MOND to galaxies accounts for their dynamics without the need for dark matter. Various scaling relations find a natural explanation in this context. This concerns in particular an algebraic relation (the Radial Acceleration Relation, RAR) between the measured acceleration and the gravitational acceleration due to the baryons alone, which has been observationally extended by orders of magnitude in acceleration using weak lensing on thousands of early- and late-type galaxies, in agreement with the universal predictions of MOND.
    \item Various formulations of MOND do exist. Current formulations are challenged by observations of galaxy clusters and by Solar System tests.
    \item Since its proposal, it has been recognised that MOND is non-relativistic and in need of an embedding in a fundamental theory. New relativistic theories have since then been proposed, some of them with a consistent cosmology, but they are still intricate and the search for a fundamental theory embedding MOND is ongoing.
\end{itemize}
\end{abstract}

\section{Introduction}
{\bf Modified Newtonian Dynamics (MOND) is the proposal by \citet{mil83a, mil83b, mil83c} that, below a certain acceleration scale $\boldsymbol{a_0}$, acting as a new fundamental constant of Nature, dynamics departs from the Newtonian expectation, offering an alternative solution to the missing mass problem for galactic systems. The core idea is that, in this regime, the equations of motion for point masses, or a continuous limit thereof, become spacetime scale-invariant, i.e. invariant under transformations $x^\mu \to \lambda x^\mu$, where $x^\mu$ are the spacetime coordinates and $\lambda$ a positive real number. In this limit, the gravitational constant $\boldsymbol{G}$ and the acceleration constant appear together as a product $\boldsymbol{G}\boldsymbol{a_0}$.} A less rigorous definition, but perhaps more intuitive rule of thumb is that the observed norm of the acceleration $a$, for accelerations $a \ll a_0$, is proportional to the \emph{square root} of the Newtonian expectation, $a \simeq \sqrt{a_0 a_{\mathrm{N} }}$ where $a_{\mathrm{N}}$ is the acceleration expected in Newtonian dynamics. Note that this raises the question of the existence of an effective absolute inertial frame with respect to which accelerations are defined, the answer to this question actually being theory-dependent. Importantly, MOND is not a theory {\it per se}, but rather a paradigm in need of embedding in a full-fledged, possibly covariant, theory, which may also address its shortcomings. Theories of MOND do exist, but they are not unique. Hence, what MOND has to say on cosmology depends on the specific relativistic version considered. It should be noted that the value of the acceleration scale $a_0$ is curiously of the order of $c^2\sqrt{\Lambda} \sim c H_0$ where $c$ is the speed of light, $\Lambda$ is the cosmological constant, and $H_0$ is the Hubble-Lema\^itre constant, which might suggest a deep connection of MOND to cosmology. Note that MOND may either be the result of a modification of gravity, or of a modification of inertia, in the sense that the inertial term $m a$ would become $m a^2/a_0$ in the MOND regime. If MOND is modified inertia it would also extend to non-gravitational forces. 

\subsection{The crux of the missing matter problem in galaxies}

Consider the outskirts of a galaxy with the approximation that the mass distribution is spherical. Assuming that gaseous tracers are on circular orbits with speed $V$ caused by the gravitational field of the galaxy, equating the centripetal acceleration to the gravitational force, $a = g$, implies, when the gravitational force is Newtonian, that $V^2/r = G M(r)/r^2.$
When most of the baryonic mass has been enclosed, $M(r)\to M$, the rotation curves should be Keplerian and falling, $V = \sqrt{GM/r}$. This is \emph{not} what is observed in rotationally-supported galaxies, but rather \emph{flat} rotation curves, the circular speed tending to a constant, $V \to V_{\infty}$. This discrepancy fully remains when correctly taking into account the geometry of a disk galaxy, and ultimately originates from the different scalings with distance of the observed centripetal acceleration ($\propto r^{-1}$) and the Newtonian gravitational force law ($\propto r^{-2}$). There are three ways to arrange for this.
\subsubsection{The dark matter proposal}
One way, the dark matter proposal, is to assume that not all the mass has been enclosed, that there is in addition a diffuse halo, the dark matter halo, arranged so that the total enclosed mass scales approximately linearly with radius in the relevant range, $M(r) \propto r$, cancelling the extra dependence of $r$ in the denominator of the gravitational force law, and hence leading to flat rotation curves. The corresponding total density should then scale as $\rho(r) \propto 1/r^2$. The current standard $\Lambda$CDM (``Lambda cold dark matter'') cosmological model relies on this dark matter proposal as one of its pillars, and has actually succeeded in explaining a large array of extragalactic and cosmological observations. In this model, numerical simulations in the absence of additional baryonic effects predict a universal density profile for dark matter halos in which galaxies should form, known as the Navarro-Frenk-White (NFW) profile, $\rho_{\mathrm{NFW}}(r) \propto \left(r/r_s\right)^{-1} (1+r/r_s)^{-2}$, with scale radius $r_s$. It has power-law scalings with slopes $-1$ (``cuspy'') centrally and $-3$ in the far outskirts, and only for intermediate $r\sim r_s$ has a $-2$ slope transition. This means that flat rotation curves should not extend indefinitely in dark matter models. More generally, since baryons still have a nonzero contribution to the rotation curve in the flat regime, the halo parameters `conspire' with the baryonic distribution in order to keep the total circular velocity constant with radius \citep{Albada}.
\subsubsection{MOND as modified gravity}
The other way to cancel the extra factor of distance in the denominator of the gravitational force law is to assume that the gravitational acceleration is the square root of the Newtonian expectation, $g \propto \sqrt{g_{\mathrm{N} }}$, with the dimensionful constant of proportionality provided by the acceleration scale $a_0$ so that $g = \sqrt{a_0 g_\mathrm{N}}$. This can only be the case at low accelerations ($g\ll a_0$), since Newtonian gravity works well at high ones. In this regime, known as the deep-MOND regime, equating the centripetal acceleration with $g$ means that the circular speed is automatically constant (flat rotation curve)
\begin{align}
\frac{V^2}{r}=\sqrt{a_0 g_\mathrm{N}} = \sqrt{\frac{GM a_0}{r^2} } \Rightarrow V^2 = \sqrt{G M a_0} = V_\infty^2.
\label{modgrav}
\end{align}
\subsubsection{MOND as modified inertia}
The third way to accomplish the same is to assume that, for low accelerations, the inertial term $m a$ changes to $m a^2/a_0$ for circular orbits, but the form of the gravitational force stays Newtonian. Then
\begin{align}
m \frac{a^2}{a_0} = m \frac{V^4}{r^2} \frac{1}{a_0} = \frac{GMm}{r^2} \Rightarrow V^4 = G M a_0 = V_\infty^4.
\label{inertialterm}
\end{align}
leading to the same conclusion. This is MOND as modified inertia. It is apparent that the above equation of motion for a test mass on a circular orbit is indeed scale-invariant under transformations $(t,r) \to \lambda (t,r)$, and the scaling of velocity with mass only involves the product $Ga_0$. This is also the case for Eq.~\eqref{modgrav}. 

\subsection{Interpolating functions}
Once considering the whole range of accelerations, and not just the Newtonian and deep-MOND limits, it is natural to consider that there should be a smooth transition from one regime to the other. There is no existing clear derivation of such a transition from first principles, and so it is approximately inferred from data. This so-called {\it interpolating function} is conjectured to arise from a more fundamental theory, but only as an approximation, just as there is no ``interpolating function'' connecting general relativity (GR) to Newtonian gravity. The interpolating function as an approximation may even vary depending on the system considered, but still have $a_0$ as a scale of transition to new dynamics.
Such an interpolating functions is denoted as $\mu(x)$ or $\nu(x)$ and relates the acceleration $a$ to the Newtonian expectation $a_\mathrm{N}$ through the algebraic relation
\begin{align}
a_{\mathrm{N}} = \mu(a/a_0) \, a
\label{eqmu}
\end{align}
or
\begin{align}
a = \nu(a_\mathrm{N}/a_0) \,a_\mathrm{N},
\label{eqnu}
\end{align}
where $\mu(x)$ and $\nu(x)$ should tend to $1$ for large (non-dimensionalised) accelerations $x \gg 1$. For low (non-dimensionalised) accelerations $x \ll 1$, one should have $\mu(x) \to x$ or $\nu(x) \to 1/\sqrt{x}$, implying the deep-MOND limit.
The mathematical relation between $\mu$ and $\nu$ is that $\nu(y) = \tilde{\mu}^{-1}(y)/y$ where $y=\tilde{\mu}(x) \equiv \mu(x)x$. The function $\tilde{\mu}(x)$ must always be a monotonically increasing function of $x$.

\section{Non-relativistic MOND theories}
An elephant in the room is that blindly applying Eq.~\eqref{eqmu} or Eq.~\eqref{eqnu} to a set of massive bodies leads to undesirable behaviours. If two bodies of different mass $m_1$ and $m_2$ were to move according to $\vec{a} = \nu(|\vec{a}_{\mathrm{N}}|) \vec{a}_{\mathrm{N}}$, the total momentum $\vec{P} = \sum m_i \vec{v}_i$ would not be conserved. Considering accelerations that are in the deep-MOND regime and the Newtonian force between them to be $\vec{F}_\mathrm{N}$ then $\dot{P} = \sqrt{a_0 F_\mathrm{N}} \left(\sqrt{m_1} - \sqrt{m_2} \right)$, which is non-zero, unless the two masses happen to coincide. This would for instance lead to the self-acceleration of the center-of-mass of a wide binary stellar system of unequal masses. Of course, precision tests of the conservation of linear momentum tests are only accessible in the high gravity regime, but another spurious effect of the blind application of such a modification would be that any system that would have high internal accelerations would see the modification of dynamics of its center-of-mass canceled, even if it were to move in a low-acceleration gravitational field. Said otherwise, the high acceleration within star clusters, or at the surface of stars themselves, would prevent clusters --- or stars themselves --- to follow a MOND orbit in the outskirts of a galaxy.
Clearly, a more careful approach is needed. This is found in the least action principle, which safeguards the theory from most pathological behaviours and guarantees usual conservation laws. There are two such prominent non-relativistic variational formulations of MOND, known as AQUAL and QUMOND, though these are not the only possible ones.

\subsection{The A-quadratic Lagrangian (AQUAL), with the peculiar power $\mathbf{3/2}$}

A non-relativistic field equation with MOND behaviour can readily be constructed \citep{BM84}. Since the Newtonian acceleration is the gradient of the Newtonian potential, $\vec{a}_{\mathrm{N}} = -\nabla \Phi_{\mathrm{N}}$, and relates to the density through the Poisson equation
\begin{align}
\nabla^2 \Phi_{\mathrm{N}} = 4 \pi G \rho ,
\end{align}
it implies that $4\pi G \rho = - \nabla \cdot \vec{a}_\mathrm{N}$. Inspired by Eq.~\eqref{eqmu}, one can now write
\begin{align}
- \nabla \cdot \left( \mu(|\vec{a}|/a_0)\,  \vec{a} \right) = 4 \pi G \rho .
\end{align}
Demanding that the acceleration should come from the gradient of a scalar, the MOND gravitational potential $\Phi$, such that $\vec{a} = -\nabla \Phi$, the MOND non-relativistic equation becomes
\begin{align}
\nabla \cdot \left( \mu(|\nabla \Phi|/a_0) \nabla \Phi \right) = 4 \pi G \rho
\label{AQUAL}
\end{align}
which, in the deep-MOND limit, gives
\begin{align}
\nabla \cdot (|\nabla \Phi| \nabla \Phi) = 4 \pi G a_0 \rho.
\end{align}
Like for solutions of the Poisson equation, the non-relativistic MOND potential is only defined up to a constant, the equation being invariant under $\Phi \to \Phi + c$ (shift symmetry). As discussed by \citet{B2007}, Eq.~\eqref{AQUAL} is analogous to Gauss' law describing the electric field inside a dielectric medium,
$\nabla \cdot (\mu_e \vec{E}) = 4 \pi \rho_e$,
where $\vec{E}$ is the electric field, $\rho_e$ the free charge density, and $\mu_e = 1 + \chi_e$ the dielectric coefficient, where $\chi_e(E)$ is the electric susceptibility that can be a function of the amplitude of the electric field in a non-linear medium. Recasting the interpolating function as $\mu(a/a_0) = 1 + \chi(a/a_0)$, one can rewrite Eq.~\eqref{AQUAL} as
\begin{align}
\nabla^2 \Phi = 4 \pi G (\rho -  \nabla \cdot \vec{\Pi}),
\label{dipolar}
\end{align}
where $\vec{\Pi} = -(\chi \vec{a})/(4 \pi G)$ is analogous to a `polarization field'\footnote{This analogy has led to the proposal of hybrid dark matter models making $-  \nabla \cdot \vec{\Pi}$ a true source of a (Newtonian) gravitational field, by considering it to be generated by a gravitational polarization vector field carried by a dark matter fluid with low rest mass density, having an internal force to stabilise the configuration \citep[e.g.,][]{Blanchet_2009,Stahl_2022}.}. In particular, the susceptibility is negative, $\chi < 0$, and so there is anti-screening, the original gravitational field being enhanced.

Eq.~\eqref{AQUAL} is actually also equivalent to writing 
\begin{align}
\mu(|\vec{a}|/a_{0}) \, \vec{a} = \vec{a}_N + \nabla \times  \vec{A},
\label{curl}
\end{align}
and is thus equivalent to Eq.~\eqref{eqmu} in vectorial form, up to a curl field $\vec{d}=\nabla \times  \vec{A}$, while it is precisely equal to it in spherical symmetry where the curl field vanishes. It is not surprising that a potential $\Phi$, whose gradient gives the acceleration $\vec{a} = - \nabla \Phi$ that satisfies the relation $\mu(|\vec{a}|/a_0) \vec{a} = \vec{a}_{N}$, cannot in general be found, but that a correction appears. Such an acceleration would indeed also have to satisfy the inverse relation $\vec{a} = \nu(|\vec{a}_{\mathrm{N}}|/a_0) \vec{a}_{\mathrm{N}}$, and if $\vec{a} = -\nabla \Phi$ comes from the gradient of a potential that means that its curl is zero and so the curl of the RHS should also be zero, which implies that $\nabla |\nabla \Phi_{\mathrm{N}}| \times \nabla \Phi_{\mathrm{N} } = 0$, which is a statement about the Newtonian configuration and is not true for general configurations. 

This formulation, first proposed by \citet{BM84}, is the epitome of a classical modified gravity (see Sect.~1.1.2) formulation of MOND. It can be derived from an action principle. For a set of massive particles with masses $m_i$, with density $\rho(\vec{x},t)=\sum_i m_i \delta({\vec{x}} - {\vec{x}}_i(t))$, where $\delta$ is the Dirac delta, and velocity field $\vec{{\rm v}}(\vec{x},t)$, the Newtonian Lagrangian density in the non-relativistic limit reads 
\begin{align}
\mathcal{L} = - \rho \left(\Phi - \frac{1}{2} \vec{\rm{v}} \cdot \vec{\rm{v}}  \right)  - \frac{1}{8 \pi G} \mathcal{L}_\mathrm{NG},
\label{lagrangian}
\end{align}
where
\begin{align}
\mathcal{L}_\mathrm{NG} = \nabla \Phi \cdot \nabla \Phi \equiv (\nabla \Phi)^2,
\end{align}
and the Euler-Lagrange equations for the gravitational field $\Phi$ (leading to the Poisson equation) read
\begin{align}
\partial_i \left( \frac{\partial \mathcal{L} }{ \partial \left( \partial_i \Phi \right) } \right) = \frac{\partial \mathcal{L} }{ \partial  \Phi },
\end{align}
where on the LHS the divergence of the variational derivative with respect to $\partial_{i}\Phi$ is being taken. Note that $\mathcal{L}_\mathrm{NG}$ as defined here does not have the dimensions of a Lagrangian density, which will also be the case, hereafter, each time a subscript will be used.  

Modifying gravity at the classical level means modifying $\mathcal{L}_\mathrm{NG}$ hereabove. In order to arrange for the differential operator in the deep-MOND regime, $\nabla \cdot \left( |\nabla \Phi|\nabla \Phi \right)$, which has two powers of $\nabla \Phi$ inside the divergence, the original Lagrangian must have three, namely $|\nabla \Phi|^3$. Written in terms of the \emph{scalar} $\nabla \Phi \cdot \nabla \Phi$, that would then be $\left( \nabla \Phi \cdot \nabla \Phi\right)^{3/2}$. This is the origin of the non-canonical kinetic term with the peculiar power $3/2$, giving its name to this classical modified gravity theory: as it is not quadratic (or a-quadratic), it is called AQUAL, as an acronym of `A-quadratic Lagrangian'. Now, letting $Y = (\nabla \Phi/a_0) \cdot (\nabla \Phi/a_0)$, in order to find an appropriate Lagrangian in the general case with $\nabla \cdot \left( \mu(|\nabla \Phi|/a_0) \nabla \Phi \right)$, one must find a function $\mathcal{F}$ such that $\mathcal{F}'(Y) = \mu(\sqrt{Y})$, as that function will be found differentiated inside the divergence operator, in other words the integral, $\mathcal{F}(Y) = \int \mathrm{d}Y\, \mu(\sqrt{Y} )$. The gravitational Lagrangian density of AQUAL replacing $\mathcal{L}_\mathrm{NG}$ is 
\begin{align}
\mathcal{L}_{\mathrm{AQUAL}} = a_0^2 \mathcal{F}((\nabla \Phi)^2/a_0^2),
\label{lagAQUAL}
\end{align}
where
\begin{align}
\mathcal{F}(Y) \rightarrow Y \; {\rm for} \; Y \gg 1 \; {\rm and} \; \mathcal{F}(Y) \rightarrow \frac{2}{3}Y^{3/2} \; {\rm for} \; Y \ll 1.
\end{align}

\subsection{Quasi-linear MOND (QUMOND)}
Another non-relativistic realisation of MOND, introduced in \citet{Milgrom_2010}, is Quasi-linear MOND (QUMOND). Unlike AQUAL, it contains two scalar fields, the MOND potential $\Phi$ and the Newtonian potential $\Phi_\mathrm{N}$. The Newtonian potential obeys the Poisson equation
\begin{align}
\nabla^2 \Phi_{\mathrm{N}} = 4 \pi G \rho
\end{align}
while a second equation enforces
\begin{align}
\nabla^2 \Phi = \nabla \cdot \left( \nu \left( |\nabla \Phi_\mathrm{N}| \right) \nabla \Phi_\mathrm{N} \right).
\end{align}
The field equation is therefore Poisson with a modified RHS depending only on the Newtonian potential, that absorbs the whole non-linearity of the theory. To get this set of equations, one now needs to replace $\mathcal{L}_{\mathrm{NG}}$ with
\begin{align}
\mathcal{L}_{\mathrm{QUMOND}} =  2 \nabla \Phi \cdot \nabla \Phi_\mathrm{N} - a_0^2 \mathcal{Q}(( \nabla \Phi_\mathrm{N})^2/a_0^2 ). 
\label{lagQUMOND}
\end{align}
where $\mathcal{Q}(Y)$ is chosen so that $\mathcal{Q}'(Y)= \nu(\sqrt{Y})$.
As in AQUAL, the gradient of the QUMOND potential $\nabla \Phi$ does not, in general, equal $\nu(|\nabla \Phi_{\mathrm{N}}|/a_0) \nabla \Phi_{\mathrm{N}}$, as $\nabla \Phi$ is only the curl-free component of $\nu \nabla \Phi_{\mathrm{N}}$. Defining the difference between them as $\vec{d^Q} = \nu \nabla \Phi_{\mathrm{N}} - \nabla \Phi$, its curl, i.e., $\nabla \times \vec{d^Q} = \nabla \nu  \times \nabla \Phi_{\mathrm{N}} = \nu' \nabla |\nabla \Phi_{\mathrm{N}}| \times \nabla \Phi_{\mathrm{N}} \equiv \vec{j}$, is a known quantity given by the Newtonian theory, and since $\vec{d^Q}$ is divergence-less the difference $\vec{d^Q}$ relates to $\vec{j}$ as if $\vec{d^Q}=\nabla \times  \vec{A^Q}$ were the magnetic field of the Biot-Savart law. This current $\vec{j}$ vanishes in spherical symmetry where all vectors are radial, and the QUMOND result then coincides with Eq.~\eqref{eqnu}. It is important to note that the general curl correction is {\it not} the same in QUMOND and AQUAL.

\subsection{Non-linearity, the External Field Effect and the Equivalence Principle}
Unlike the Poisson equation, the MOND field equation is a non-linear equation, which mathematically in the case of AQUAL (in deep-MOND) is governed on the LHS by the $p$-Laplacian $\nabla \cdot (|\nabla \Phi|^{p-2} \nabla \Phi )$ with $p=3$.  
As it is non-linear, the total potential is not the simple sum of individual potentials in the low acceleration regime $a \lesssim a_0$. This gives rise to the External Field Effect (EFE). Motion within a stellar system that should have been in deep-MOND is not so if this system is immersed in a stronger gravitational field. 
Considering an internal gravitational field, $\Phi_{\mathrm{int} }$, subject to a linear external potential, i.e., constant acceleration, $\vec{a}_\mathrm{ext}$, and inserting the total field $-\nabla \Phi = \vec{a}_{\mathrm{ext} } - \nabla {\Phi}_{\mathrm{int}}$ into the AQUAL equation in the deep-MOND limit (assuming here $a_0 > |a_\mathrm{ext}|$ in a galactic situation) leads to
\begin{align}
\nabla \cdot \left(|\nabla \Phi_{\mathrm{int}}  - \vec{a}_\mathrm{ext} | \left( \nabla \Phi_{\mathrm{int}} - \vec{a}_\mathrm{ext} \right) \right) = 4 \pi G a_0 \rho
\end{align}
With the divergence of a constant $\nabla \cdot \vec{a}_\mathrm{ext}$ being zero, and assuming that the external force is dominant, $|\vec{a}_{\mathrm{ext}}| \gg |\nabla \Phi_{\mathrm{int}}|$, which shall be assumed henceforth, leads to
$\nabla \cdot \left(\nabla \Phi_\mathrm{int} \right) = 4 \pi G a_0 \rho/| \vec{a}_\mathrm{ext} | $
which implies that the internal field in this limit behaves as in Newtonian gravity, \emph{but} with a gravitational constant $G$ that is larger $G \to G a_0/ |\vec{a}_\mathrm{ext}|$. This zeroth order approximation has however ignored the inherently anisotropic character of the effect, singled out by the direction of the external force. While the primary effect is to renormalise $G$ internally, the secondary effect turns out to be a stretching of the gravitational field, i.e., its equipotential surfaces, in the direction of the external field. Assuming that the gravitational field tends to the constant external field asymptotically, linearising the QUMOND or AQUAL equation, it is found that $\Phi_{\mathrm{int}}(r)$ goes to the Newtonian $\Phi_{\mathrm{N},\mathrm{int}}(r)$, \emph{but} with the replacements $G\to G/\mu$ and $r\to (1+ \log'(\mu) \sin^2(\theta) )^{1/2} r $ in AQUAL, and $G \to G \nu$ and $r \to r/(1+ \log'(\nu) \sin^2(\theta )/2)$ in QUMOND, where $\mu$, $\nu$ and their logarithmic derivatives are evaluated at $|\vec{a}_{\mathrm{ext}}|/a_0$, and $\theta$ is the angle with respect to the external field direction.  Note that the correction is a quadrupole ($m=2)$ due to the fact that $\vec{a}_{\mathrm{ext}} \gg \nabla \Phi_{\mathrm{int}}$. When both are of the same order of magnitude, a dipolar ($m=1$) effect also arises, which can be computed only numerically (see Sect.~3.6).

The EFE implies that the Strong Equivalence Principle is broken as it would not be possible to cancel all gravitational effects by moving to a freely falling frame. Then it will matter \emph{where} the motion happens with respect to the external gravitational field. The internal dynamics of a satellite galaxy will for instance depend on where it lies with respect to its host \citep{McMil}. As GR emblematically obeys both the Strong and Weak Equivalence Principle, it is worth taking as a reference with which to contrast. Orbits of test masses are geodesics of the metric $g_{\mu \nu}$ with no reference to their composition, thus satisfying the Weak Equivalence Principle. In GR it is additionally possible to find a frame, the free-falling frame, in which, locally, $g_{\mu \nu} \to \eta_{\mu \nu}$, the Minkowski metric, thus cancelling all gravitational interactions. This does not however suffice in extensions of GR where there will typically be a scalar field $\phi$ or vector field $A_{\mu}$ that it is not possible to remove (see Sect.~4). As a massive body will typically source the scalar field or vector field, geodesics will be affected depending on whether those are near or far away from the source by their effect on the metric. Scalar-tensor theories, for instance, generically break the Strong Equivalence Principle (as does almost every extension of GR).

\subsection{Phantom densities and numerical solvers}
It can be useful to think in Newtonian terms and consider the force that differs from the Newtonian setting to arise due to a fictitious mass distribution obeying Newtonian gravity. This is called the phantom mass (density). Depending on the configuration it can be negative in certain regions (which has been suggested as a distinct signature of MOND).

Thinking in Newtonian terms amounts to imposing the Poisson equation on $\Phi$, namely $\nabla^2  \Phi = 4 \pi G \rho_{\mathrm{tot}}$ where $\rho_\mathrm{tot}$ is the total mass density in Newtonian gravity. Baryons are then subtracted from this total (or dynamical Newtonian) density to reveal the phantom density by subtracting $\nabla^2 \Phi_{\mathrm{N}}$. 
In AQUAL, the MOND solution $\Phi$ must first be found and then the phantom density can be derived by subtracting $4 \pi G \rho = \nabla \cdot \left( \mu \left(|\nabla \Phi|/a_0\right) \nabla \Phi \right)$ from $\nabla^2 \Phi$ so that 
\begin{align}
\rho^{\mathrm{AQUAL}}_{\mathrm{ph}} = -  \frac{\nabla \cdot \left( \chi(|\nabla \Phi|/a_0) \nabla \Phi \right)}{4 \pi G}
\end{align}
where $\chi \equiv \mu -1$. In QUMOND, on the other hand, $\nabla^2 \Phi$ is directly related to the Newtonian gravitational field $\nabla \Phi_{\mathrm{N}}$, so the expression for the phantom density $\rho_{\mathrm{ph}}$ is 
\begin{align}
\rho^\mathrm{QUMOND}_{\mathrm{ph}} = \frac{\nabla \cdot \left( \overline{\nu}\left(|\nabla \Phi_{\mathrm{N}}|/a_0\right) \nabla \Phi_{\mathrm{N}} \right)}{4 \pi G} 
\end{align}
where $\overline{\nu} \equiv \nu -1$. Hence, it is not necessary to have the MOND solution to know the phantom distribution in QUMOND, as it is determined by the Newtonian field. This makes it particularly convenient to devise numerical Poisson solvers in QUMOND, first getting the Newtonian potential, then computing the phantom density on a grid, and finally solving the Newtonian Poisson equation a second time with appropriate boundary conditions and the phantom density as a source. This is the basis of the Poisson solver of the publicly available \textsc{Phantom of Ramses} patch \citep[\textsc{por} patch,][]{Lughausen_2015, Nagesh_2021} of the adaptive mesh refinement (AMR) N-body/hydrodynamical code \textsc{ramses}, numerically implementing the QUMOND Poisson equation with the above strategy.

\subsection{Some general relations and exact solutions}
Here are some general relations and exact solutions to the MOND field equations, valid both in AQUAL and QUMOND.

\subsubsection{The vacuum solution}
In vacuum, away from a point source located at $r=0$, the deep-MOND equation in spherical coordinates reduces to $(r^2 (\Phi'(r))^2 )'/r^2 = 0$, which can readily be solved giving $\Phi(r) = c_2 \log(r/r_0)$, which, when matched to the requirement that $ a = \sqrt{a_0 a_{\mathrm{N} }}$ gives that $c_2 = \sqrt{GMa_0}$ and so that $\Phi(r) = \sqrt{GM a_0} \log(r/r_0) $. More generally, $\Phi(r) = \int^{r} \mathrm{d}r'\, \nu(|\nabla \Phi_{\mathrm{N}}|) \nabla \Phi_{\mathrm{N}}$. While it suffices to provide a logarithmic potential in order to explain flat rotation curves at large radii, the mass-dependent scaling $\sqrt{M}$ of its strength is actually the key to the MOND phenomenology.

In the deep-MOND regime the phantom density $\rho_{\mathrm{ph}}$, equal in QUMOND and AQUAL due to spherical symmetry, is, $\rho_{\mathrm{ph}} = M/[ 4\pi r^2 r_M]$, where $r_M = \sqrt{GM/a_0}$, giving the right mass distribution that leads to constant rotation curves. It is seen to be centred on the mass, but \emph{diffuse}. It would be a misconception to think that MOND predicts modifications to the gravitational potential that track the distribution of ordinary matter, as this most simple example illustrates. 

\subsubsection{The virial relations and the two-body force}

The scalar virial theorem states that for a time-independent system in any gravitational theory, $2 K + W = 0$, where $K$ is the total kinetic energy of the system and $W= - \int \rho {\vec{x}} \, \cdot \nabla \Phi \, d^3x$ is proportional to the total potential energy. In the deep-MOND regime for an ensemble of test-particles with individual masses much smaller than the total mass of the system $M$, one has that $W=(-2/3)\sqrt{GM^3a_0}$, so that the mass-weighted mean squared velocity $\langle {\rm v}^2 \rangle = (2/3) \, \sqrt{GMa_0}$. Considering that the observed line-of-sight velocity dispersion $\sigma$ of a system such as an isolated dwarf galaxy in the deep-MOND regime would approximately be (and exactly in the isotropic case) $\sigma^2 = \langle {\rm v}^2 \rangle/3$, one gets
\begin{align}
\sigma^4 = \frac{4}{81} GMa_0 .
\label{eqsigmavir}
\end{align}
If one considers instead a system of several non-test mass bodies, whose extents are much smaller than the separations between them, one has that the deep-MOND point mass virial relation is $\sum_i \vec{x}_i \cdot \vec{F}_i = (-2/3) \sqrt{GM^3a_0} (1 - \sum_i (m_i/M)^{3/2})$. This yields a very general relation for the mass-weighted mean squared velocity $\langle {\rm v}^2 \rangle$ of such a system in the deep-MOND regime:
\begin{align}
\langle {\rm v}^2 \rangle = \frac{2}{3} \sqrt{GMa_0} \left(1 - \sum_i \left( \frac{m_i}{M} \right)^{3/2}\right).
\end{align}
Restricting the expression of the point mass virial relation to a system of two bodies also yields the general form of the two-body force between two bodies of masses $m_1$ and $m_2$ separated by a mutual distance $r$ in the deep-MOND regime:
\begin{align}
F_{\rm 2body} = \frac{2}{3} \left[ (m_1+m_2)^{3/2} - m_1^{3/2} - m_2^{3/2} \right] \frac{\sqrt{Ga_0}}{r}.
\end{align}
Interestingly, this means that the acceleration felt by a body does depend on its mass. However, Taylor expanding the expression in the limit of $m_1 \ll m_2$ yields $F_{\rm 2body} \to m_1 \sqrt{Gm_2a_0}/r$, meaning that a test-mass $m_1$ orbiting a massive body $m_2$ does indeed feel the usual deep-MOND acceleration $\sqrt{a_{\rm N} a_0}$. These virial relations and the expression of the two-body force in the deep-MOND limit actually remain valid in any formulation of MOND as modified gravity \citep{Milgrom_2014b}.

\subsubsection{The isothermal sphere and polytropes}
Consider a gas or an isotropic pressure-supported stellar system with constant temperature $T$ or constant velocity dispersion $\sigma$, under the influence of its gravity.
The condition of hydrostatic equilibrium, or the Jeans equation in the absence of velocity anisotropies, is $\sigma^2 \mathrm{d} \log \rho / \mathrm{d} \log r =-rg$, where $\sigma^2 = k_B T / m$ and $g$ is the acceleration due to gravity. If $\mathrm{d} \log \rho / \mathrm{d} \log r$ is constant, it immediately follows that the gravitational acceleration $g$ must go like $r^{-1}$, which in Newtonian gravity leads to $M(r) \propto r$ and hence a $-2$ power-law slope for the density. The singular isothermal sphere is a particular solution to the Newtonian case that assumes a single power-law slope for the density everywhere. Inserting $g = a_{\mathrm{N}} = GM(r)/r^2$, and assuming that $\rho(r)= A r^{-\alpha}$ then gives $- \alpha \sigma^2 = -4 \pi G A r^{2-\alpha}/(3-\alpha)$ with a clear solution at $\alpha=2$ and $A = \sigma^2/(2\pi G)$. In Newtonian gravity, it has the shape to lead to constant rotation curves with circular velocity $V = \sqrt{2} \sigma$. Its total mass is divergent. It is however not possible to simply turn such a solution into a MOND solution as the systems are self-gravitating and the law of gravity in each is different. Hence the equilibrium matter distributions will be different. It will still be true that $a = \nu(|a_{\mathrm{N} }|/a_0) a_{\mathrm{N}}$ in spherical symmetry, but both sides must be evaluated keeping the matter distribution fixed. 

In MOND, the gravitational acceleration will go like $r^{-1}$ outside the bulk of the mass distribution for any finite mass. For a system fully in the deep-MOND regime, Eq.~\eqref{eqsigmavir} can be used to directly find a particular solution for a finite mass isothermal sphere. Inserting $\sigma^2 = (2/9) \sqrt{GMa_0}$ on the LHS and $g = \sqrt{GMa_0}/r$ on the RHS of the hydrostatic equlibrium condition $\sigma^2 \mathrm{d} \log \rho / \mathrm{d} \log r =-rg$  leads to $\mathrm{d} \log \rho / \mathrm{d} \log r = -9/2$.
Therefore, the density profile of the isotropic deep-MOND isothermal sphere is a power law with exponent $-9/2$. Notably, the mass is finite in MOND and proportional to the velocity dispersion to the 4th power (or temperature squared). Another  particular solution, which this time is cored, is $\rho(r) = [243\sigma^4/(16 \pi G a_0 r_0^3)] \times [(r/r_0)^{3/2}+1]^{-3}$, which when $r/r_0\to 0$ asymptotes to a constant.

In Newtonian gravity, exact solutions also exist for polytropes, for which it holds that pressure $P \propto \rho^{\gamma}$, or equivalently with the equation of state parameter $P/\rho \propto \rho^{1/n}$ where $n \equiv 1/(\gamma -1)$. The isothermal case above corresponds to $\gamma=1$. In Newtonian gravity, hydrostatic equilibrium implies the Lane-Emden equation and there are in addition analytic solutions for $n=0,1,5$. The Newtonian polytropes have finite mass for only $n \leq 5$. In MOND, other than the isothermal case, an analytic solution is known only for the case $n=0$ which is the constant-density sphere. For the constant density sphere, the Newtonian case is already known having $a_{\mathrm{N} } \propto r$, and so the deep-MOND limit has $a \propto \sqrt{r}$ inside the sphere. Some general properties are nevertheless known for deep-MOND polytropes \citep{Milgrom_polytrope}. They all have finite mass, and for $\gamma>1$ they also have finite radius.

\subsubsection{Exact solutions for disks}

As established in Sect.~2.1, Eq.~\eqref{eqmu} is exact in AQUAL when $\nabla |\nabla \Phi_{\mathrm{N}}| \times \nabla \Phi_{\mathrm{N} } = 0$, which is also true in QUMOND. This condition is always satisfied in spherical symmetry, but is not restricted to it. An example of a non-spherical system satisfying this condition outside the plane of symmetry is the Kuzmin disk, an infinitesimally thin disk with density (in cylindrical coordinates $R$ and $z$), $\rho_\mathrm{K}(R,z)= \Sigma_\mathrm{K}(R)\delta(z)$, with surface density $\Sigma_\mathrm{K}(R) = Mh/[2\pi (R^2 + h^2)^{3/2}]$ and Newtonian gravitational potential $\Phi_{\mathrm{K},\mathrm{N}} = - GM/[R^2+(|z|+h)^2]^{1/2}$. When only considering the gravitational field above or below the disk, the Kuzmin potential can also be seen as the one due to a point source of mass $M$, instead, respectively lowered or raised by $h$ above the center of the disk. As the gravitational field of a point mass is spherically symmetric, the MOND acceleration outside the disk is unambiguous (the same in QUMOND and AQUAL), and given by the vectorial version of Eq.~\eqref{eqnu} with $\vec{a}_\mathrm{N}(R,z) = GM (\vec{r}\pm \vec{h})/|\vec{r} \pm \vec{h}|^3$, where $\vec{r}$ is the radial vector in spherical coordinates and $\vec{h}$ points from the equivalent point source below the disk to the origin (the case with the positive or negative term being used when above or below the disk, respectively). This is also the sum of the radial and vertical Newtonian accelerations. In the deep-MOND limit it is possible to integrate the gradient of the potential to give $\Phi_{\mathrm{K}}(R,z) = \sqrt{GM a_0} \log(R^2 + [|z|+h]^2)/2$. The rotation velocity of this deep-MOND Kuzmin disk is then given by $V^2 = \sqrt{GMa_0} R^2/(R^2+h^2)$. Note that this exact solution {\it differs} in the plane from Eq.~\eqref{eqmu} or Eq.~\eqref{eqnu}, which would have given a rotation velocity $V^2 = R \sqrt{a_n a_0} = \sqrt{GMa_0} R^{3/2}/(R^2+h^2)^{3/4}$. Indeed, exactly in the plane of the disk, due to symmetry, the Newtonian acceleration has no vertical component, but only a radial component $a_{N,r}$. Right outside, the amplitude of the vertical acceleration is ${a}_{N,z} = 2 \pi G \Sigma$ and so the total acceleration outside is $a^{+}_{N} = \sqrt{a_{N,r}^2 + (2 \pi G \Sigma)^2}$. Hence, it is exactly true for the Kuzmin disk that the radial acceleration in the plane is $a_r = \nu(a^{+}_{N}/a_0) a_{N,r} $ in QUMOND (or in AQUAL when inverting $\mu$), where it can be seen that the \emph{vertical} component of the acceleration, directly related to the mass surface density, matters for the \emph{radial} acceleration \citep{Brada}.

\subsubsection{Approximate formula for disks}

The exact solution hereabove for Kuzmin disks can serve as a guide for establishing an approximation to the radial acceleration within general galaxy disks, either in QUMOND or in AQUAL. When Eq.~\eqref{eqmu} and Eq.~\eqref{eqnu} hold exactly outside of a razor-thin disk, it has been shown that the Newtonian argument of the $\nu$-interpolating function is simply that just outside of the disk, namely $a_N^+/a_0$ where 
\begin{align}
a_N^+ \equiv \left(a_N^2 + (2\pi G\Sigma)^2\right)^{1/2}
\end{align}
for a local baryonic disk surface density $\Sigma$, including the vertical component of the Newtonian acceleration. Then, in QUMOND or AQUAL, one has for the radial acceleration within the disk
\begin{align}
  a = \nu(a_N^+/a_0) \,a_\mathrm{N} \; \; \; \; \; \; \; \; {\rm or} \; \; \; \; \; \; \; \; a = \frac{a_N}{\mu\left(\frac{a_N^+}{a_0} \nu\left(\frac{a_N^+}{a_0}\right)\right)}.
\label{aqualRC}
\end{align}
\citet{Brada} showed for instance how good such an approximation was, in AQUAL, for computing the rotation curves of exponential disks. This approximation however ignores the subtle differences between the curl field of QUMOND and AQUAL.

\subsection{Further MOND generalisations: EMOND, GQUMOND, TriMOND}

The heart of the MOND paradigm is that the determining factor for the appearance of missing matter is acceleration and not, for instance, the mass or size of the system. However, this postulate is not {\it a priori} theoretically incompatible with the existence of other relevant scales in a generalized theory of gravity. One such generalization is the `Extended MOND' (EMOND) framework proposed in \citet{Zhao}, where the Lagrangian density of Eq.~\eqref{lagAQUAL} is replaced by
\begin{align}
    \mathcal{L}_{\mathrm{EMOND}} = \Lambda(\Phi) \, \mathcal{F}((\nabla \Phi)^2/\Lambda(\Phi)),
\end{align}
where $\Lambda$ plays the role of $a_0^2$, but which now depends on the depth of the potential well itself (with $\Lambda(\Phi) \to constant$ for small $\Phi$), which can be generalised to a naïve relativistic version in a scalar-tensor theory (à-la Sect.~4.2). Strictly, EMOND is no longer MOND in the sense of spacetime scale invariance, and MOND is only recovered in the limit of shallow potentials. It involves a new scale beyond $a_0$, but this scale can be chosen so that only the Hubble-Lemaître constant or the cosmological constant effectively appears in the Lagrangian. 
This would in principle allow for a larger effective value of $a_0$ in the center of galaxy clusters, which might be helpful to explain observations (see Sect.~3.4) but would also imply a different $a_0$ scaling for galaxies residing inside clusters, for which there is no evidence \citep[although see][for possible hints of such an $a_0$ rescaling]{Freundlich}.

The Lagrangian density of Eq.~\eqref{lagQUMOND} can also be generalized to higher derivatives of the auxiliary potential
\begin{align}
\mathcal{L}_{\mathrm{GQUMOND}} =  2 \nabla \Phi \cdot \nabla \Phi_\mathrm{N} - a_0^2 \mathcal{P}(\Phi_\mathrm{N}, \nabla \Phi_\mathrm{N}, \nabla^2 \Phi_\mathrm{N},...,\nabla^n \Phi_\mathrm{N}). 
\end{align}
Such a `Generalized QUMOND' (GQUMOND) theory \citep{Milgrom_2023a} typically allows for the introduction of additional dimensional constants besides $a_0$, such as a characteristic length-scale or a frequency-scale, beyond which MOND effects can be suppressed or screened, which could be helpful to pass Solar System constraints (see Sect.~3.5).

One can also generalize the Lagrangian density of Eq.~\eqref{lagQUMOND} by considering two auxiliary potentials, $\Phi_{\rm N}$ and $\varphi$, instead of just one, namely
\begin{align}
\mathcal{L}_{\mathrm{TriMOND}} =  2 \nabla \Phi \cdot \nabla \Phi_\mathrm{N} - a_0^2 \mathcal{P}((\nabla \Phi_\mathrm{N})^2/a_0^2, (\nabla \varphi)^2/a_0^2, (2 \nabla \Phi_\mathrm{N} \cdot \nabla \varphi)/a_0^2). 
\end{align}
Such a `Tripotential MOND' (TriMOND) theory \citep{Milgrom_2023b} can generate a rich variety of predictions beyond those of AQUAL and QUMOND in non-symmetric situations, even in the deep-MOND regime which would now be described by a function of two variables.

\subsection{MOND as modified inertia}

To account for special relativity, the first term (depending on the velocity) of the Lagrangian of Eq.~\eqref{lagrangian} for a single particle moving at speed ${\rm v}$ actually has to be ``modified'' by a factor $\gamma({\rm v})=1/\sqrt{1-({\rm v}/c)^2}$). One could therefore imagine that something similar happens for the MOND regime of low acceleration, where $\mu(a)$ would play a role similar to $\gamma({\rm v})$ in special relativity, to obtain MOND as ``modified inertia'' (Sect.~1.1.3). No full-fledged theory of MOND as modified inertia  exists, but \citet{Milgrom_1994} has shown that doing so while keeping the standard symmetries (space and time translations, rotation, and Galilean symmetry) implies a time-nonlocal theory (noting that one could nevertheless also imagine a non-relativistic boost symmetry other than Galilean and specific to MOND). In such time-nonlocal theories, the inertial term in the particle equation of motion becomes $m \mathcal{A}[\{{\vec{x}}(t)\}, a_0]$, where $\mathcal{A}$ is a (vector) functional of the whole trajectory of the particle $\{{\vec{x}}(t)\}$, with dimension of acceleration. The formal Newtonian and deep-MOND limits of such a theory should correspond, respectively, to $a_0 \rightarrow 0, \mathcal{A} \rightarrow {\rm d}^2\vec{x}/{\rm d}t^2$ for Newton, and $a_0 \rightarrow \infty, \mathcal{A}[\{\vec{x}(t)\}, a_0] \rightarrow \mathcal{Q}[\{{\vec{x}}(t)\}]/a_0$ where $\mathcal{Q}$  is a (vector) functional with dimension of acceleration squared, for the deep-MOND regime. It is possible \citep{Milgrom_1994} to design toy-models for which $\mathcal{Q}[\{{\vec{x}}(t)\}] = -a^2 \vec{1}_r$ for circular orbits, thereby giving back Eq.~\eqref{inertialterm}. Various toy models of this type have been described and applied to non-relativistic N-body systems in \citet{Milgrom_mi}. Such models actually differ from modified gravity formulations in various secondary MOND predictions, such as the EFE that can depend, for instance, on the frequency ratio of the external and internal field variations. It is also important to keep in mind that Eq.~\eqref{eqmu} and Eq.~\eqref{eqnu} can be exactly valid for circular orbits inside a flattened disk only in such modified inertia formulations.

\section{Phenomenology}

In this Section, the observationally successful predictions of MOND on galaxy scales are briefly presented, as well as some of its most salient observational challenges. These are mostly independent of the exact formulation of MOND chosen, and rely (in most cases) on the simple rule of Eq.~\eqref{eqmu} or Eq.~\eqref{eqnu}.

\subsection{The BTFR of disk galaxies}

\begin{figure}[t]
\centering
\includegraphics[width=0.8\textwidth]{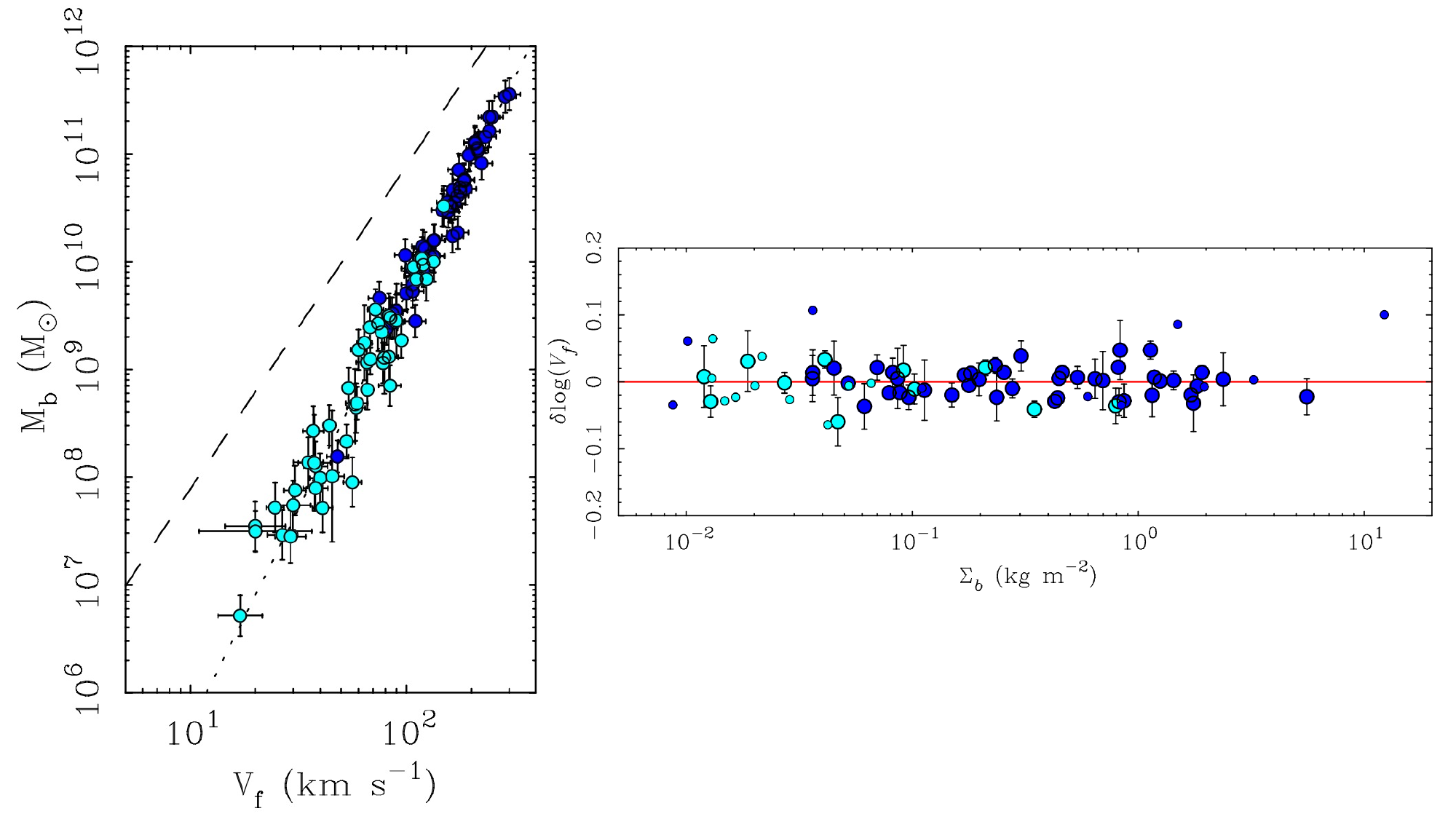}
\vspace{-0.1cm}
\caption{Left panel : the baryonic Tully-Fisher relation (BTFR) from \citet{Famaey_2012}. The dark blue points are galaxies with a larger stellar mass than observed gas mass, while the light blue points are the opposite. The dotted line is the MOND prediction, the dashed line represents the relation one would have from a universal cosmic baryon fraction in $\Lambda$CDM. Right panel: residuals from the BTFR as a function of the characteristic baryonic surface density. Galaxies with uncertainty on the rotation velocity less than 5\% are shown as larger points. }
\label{fig1}
\end{figure}

As it readily appears in Eqs.~\eqref{modgrav} and \eqref{inertialterm}, a core prediction of MOND is that galaxy rotation curves should be asymptotically flat (at least until the external gravitational field dominates over the internal one) and that the asymptotic circular speed to the fourth power should be directly proportional to the baryonic mass of the galaxy. The overall amplitude of the rotation curve at a given baryonic mass should be directly proportional to $Ga_0$, which observationally implies a value $a_0 \simeq 10^{-10} {\rm m \, s}^{-2}$. This value of $a_0$ in turn puts the outskirts of observed galaxy rotation curves in the weak-acceleration regime (at least for galaxies at low redshifts, the regime not being always observationally reached at larger redshifts), far from the bulk of their baryonic mass, making the prediction internally consistent. Importantly, within the MOND paradigm, once fixing $a_0$, one {\it simultaneously} fixes both the overall amplitude of rotation curves at a given baryonic mass {\it and} the transition acceleration where missing matter effects should appear. This of course need not be the case in general, nor in an unspecified modified gravity framework, and even less so in the dark matter framework. This MOND relation $V^4 = GMa_0$ involves the {\it asymptotic} circular speed and is predicted by MOND to be a fundamental relation, implying that the observed scatter should be minimal and largely attributable to observational errors and minor geometrical effects. Such a prediction is counter-intuitive within the dark matter paradigm. Indeed, if the fraction of baryons within galaxy-sized dark matter halos were universal, one might expect a rough relation between the circular speed cubed and the baryonic mass, as such a relation is anticipated between the circular velocity at the halo virial radius and the virial mass. However, a universal baryon fraction within dark matter halos --- which was a standard assumption in the early days of $\Lambda$CDM --- would itself pose significant challenges for the $\Lambda$CDM model, as the observed luminosity function of galaxies differs markedly from the predicted halo mass function. Reconciling the observed luminosity function with the theoretical halo mass function through what is called `abundance matching' yields a stellar-to-halo mass relation where low-mass galaxies are more dark matter-dominated than Milky Way-like ones. Along with the observed scaling relation between gas and stellar masses of galaxies, this returns a relation between asymptotic circular speed and baryonic mass with a logarithmic slope of about 4 rather than 3 \citep{DiCintio}. Nevertheless, the stellar-to-halo mass relation from abundance matching would imply a turnover at high mass, and the scatter around the relation should not be negligible in $\Lambda$CDM.

This MOND-predicted relation is actually observed (see Fig.~\ref{fig1}) and known as the {\it baryonic Tully-Fisher relation} \citep[BTFR,][]{McGaugh}: it is one of the tightest astrophysical relations known, with a very small intrinsic orthogonal scatter of only $\sim 6$\%, smaller than {\it a priori} expected in $\Lambda$CDM \citep{Desmond1}. Interestingly, the scatter around the relation actually increases when using other characteristic velocities than the asymptotic one, such as the maximum velocity, which strengthens the original MOND prediction. This kinematically-deduced BTFR follows an almost perfect power-law across a five-order-of-magnitude range in mass. Interestingly, there has been a long history of false alarms \citep[see, e.g.,][]{fam2} suggesting that the BTFR might break down at low or high masses for some systems, but careful re-analyses have invariably indicated overlooked systematics \citep{lelli24}, or have substantially revised the reduced data themselves, re-aligning with the BTFR. The case of massive super-spiral galaxies, initially reported as inconsistent with the BTFR, was strikingly corrected after a re-analysis of the data \citep{diteod}. This means that there is, in fact, no change in the BTFR slope for the most massive spirals, exactly as predicted in MOND. Perhaps the most salient and under-appreciated prediction of MOND regarding the BTFR is that its (small) residuals should {\it not} depend on any secondary parameters, such as disk size or baryonic surface density at a fixed baryonic mass, nor with the residuals of the mass-size relation. As illustrated on Fig.~\ref{fig1} (right panel), this is indeed what is observed: the residuals do not depend on the baryonic surface density. This is in fact challenging to understand within the dark matter context: assuming a fixed factor between total and baryonic mass at a given baryonic mass $M$, one would expect $V^4 \sim M^2/R^2 \sim M \, \Sigma$, where $\Sigma$ is the surface density. However, observations only show $V^4 \sim M$, with no dependence on $\Sigma$ or on size. Explaining this in the dark matter context might require an anti-correlation between dark matter halo spin and concentration, assuming disks settle in the center of halos and retain a fraction of the halo's angular momentum, though the situation is likely more complex in a $\Lambda$CDM galaxy formation context. The observed BTFR, without any dependence on secondary parameters, is on the other hand perfectly in line with the original (and non-trivial) prediction of MOND.

\subsection{The diversity of rotation curves and the CSDR}

\begin{figure}[t]
\centering
\includegraphics[width=0.6\textwidth]{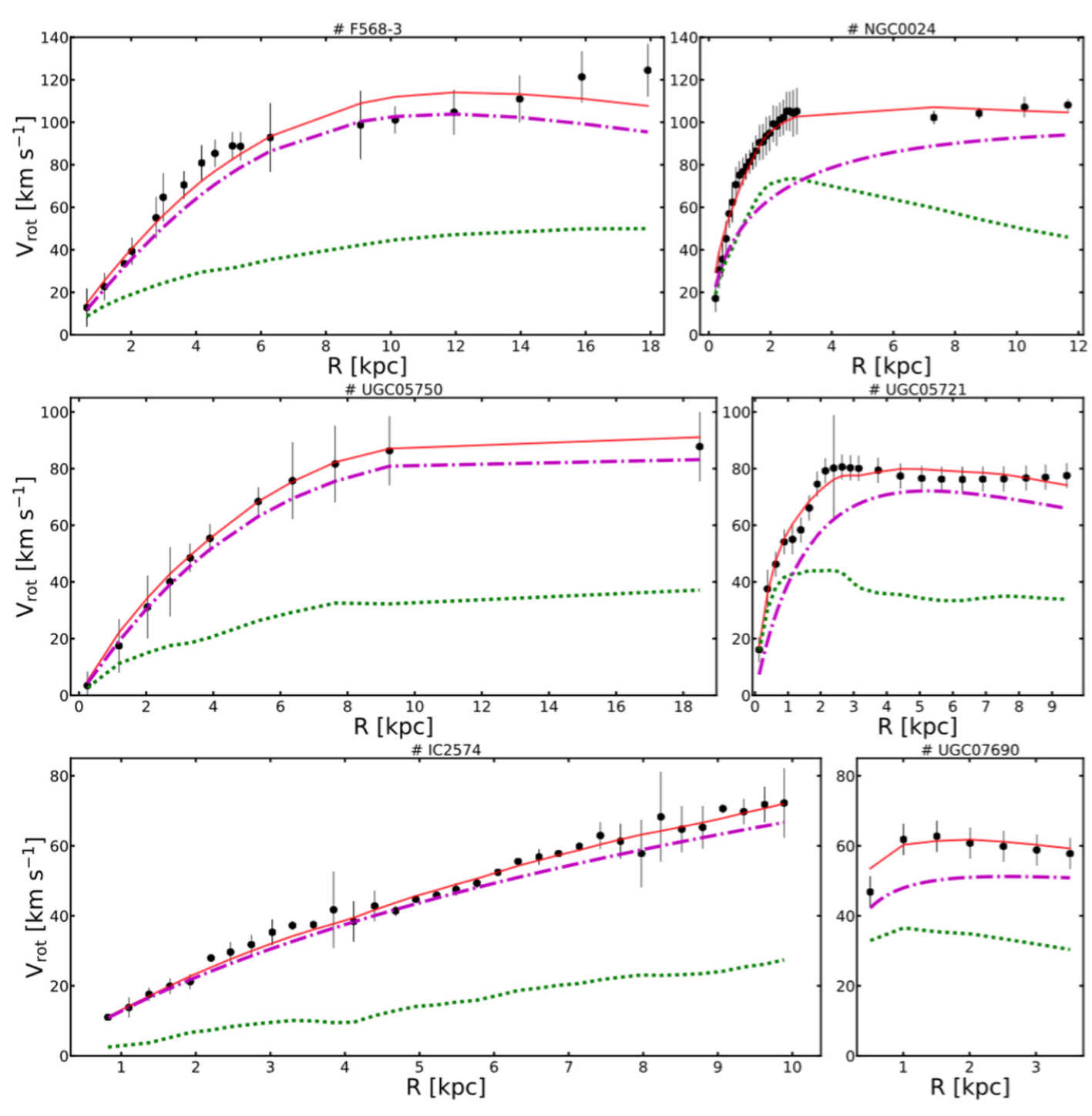}
\caption{Three pairs of (nearly) ``twin galaxies'' on the BTFR. The rotation curves are fitted (in Newtonian gravity) with Einasto profiles for the dark halo \citep[from][]{Ghari_2019}. The dotted green line is the rotation curve generated by baryons alone in Newtonian gravity, the dashed magenta line is the dark halo rotation curve, and the red curve is their combination in quadrature. This illustrates the diversity of rotation curve shapes at a given mass scale and the correlation of their shapes with the Newtonian baryonic rotation curve (dotted green line).}
\label{fig2}
\end{figure}

The truly most salient phenomenological prediction of MOND is that, whilst the BTFR should be independent of baryonic surface density, the inner shape of rotation curves should, on the other hand, depend on it, even when these rotation curves would appear to be dominated by ``dark matter'' everywhere. Consider two low-surface brightness (LSB) exponential disks, with surface density profiles $\Sigma_i(R) = \Sigma_{0i} \, {\rm exp}(-R/R_{di})$, where $\Sigma_{0i}$ is the central surface density and $R_{di}$ the exponential scale length, sharing the same total baryonic mass $M = 2 \pi \Sigma_{0i} R_{di}^2$. If $\Sigma_{0} \ll a_0/(2 \pi G)$, both disks reside in the deep-MOND regime and should appear ``dark matter-dominated'' down to their center. Their enclosed masses in scale-length units, $M(R/R_{d})$, share the exact same profile and, at a fixed number of scale-lengths, their respective squared circular velocities generated by baryons in Newtonian gravity scale as the inverse of scale-length, $V_{N1}^2/V_{N2}^2 \sim R_{d2}/R_{d1}$. Since the MOND gravity enhancement factors to the $V_{Ni}^2$ are $\nu_i=\sqrt{a_0/a_{Ni}}$, the respective enhancement factors obey $\nu_1/\nu_2 \sim R_{d1}/R_{d2}$, and the two MOND rotation curves therefore will be exactly the {\it same} when expressed in scale-length units. Hence, the lowest surface density disk, with the largest scale length, will simply display a stretched-out (by scale-length) version of the other rotation curve, with a slower rise (in physical distance units) towards the same asymptotic circular velocity. Real disk galaxies are never such pure exponential disks, but the prediction holds in general, as it immediately follows from the scale invariance of the deep-MOND limit. It turns out that this uncanny behaviour {\it is} what is actually observed. Firstly, LSB galaxies indeed appear dark matter-dominated down to their center while sharing the same BTFR as high surface brightness (HSB) ones \citep{deBlok_1997}, namely one of the core predictions of \citet{mil83b} before such LSB galaxies were actually known to exist. Moreover, the shape of the central distribution of matter and the inner rise of the rotation curve {\it are} related, as if {\it the luminous mass dominates the gravitational potential in the central regions, even in LSB dwarf galaxies} \citep{Swaters_2009}, thereby following the MOND prediction. Interpreted in the dark matter context, this would mean that ``BTFR twin'' galaxies (i.e., sharing the same baryonic mass and asymptotic circular velocity) must display a variety of inner dark matter halo profiles as a function of the surface density of the baryons, as illustrated on Fig.~\ref{fig2}. This remains very surprising today in the standard $\Lambda$CDM context, where baryonic feedback must transform the central cuspy (power-law slope of $-1$) NFW profiles into cored ones (power-law slope of $0$) in a fine-tuned fashion. Current simulations either produce too many cuspy halos when feedback is insufficient, or too many cores when feedback is overly efficient~\citep{Ghari_2019}. This modern observational puzzle in the dark matter context was however one of the core predictions originally made by \citet{mil83a,mil83b}. The prediction can be made more quantitative by considering the relation between the central baryonic surface density $\Sigma^0_{\rm b}$ of the disc and the dynamically measured face-on (integrated over the vertical direction) central surface density in Newtonian gravity, $\Sigma^0_{\rm dyn}$. This {\it Central (Surface) Densities Relation} (CSDR, or simply CDR) is predicted to asymptote to $\Sigma^0_{\rm dyn}=\Sigma^0_{\rm b}$ for $\Sigma^0_{\rm b} \gg a_0/(2 \pi G)$, and to $\Sigma^0_{\rm dyn} \propto \sqrt{\Sigma^0_{\rm b}a_0/(2\pi G)}$ for $\Sigma^0_{\rm b} \ll a_0/(2 \pi G)$, as observed. The role played by $a_0$ in this relation is very different from the one it plays in the BTFR, yet the two observed relations imply the same value. A general discussion on the CSDR can be found in \citet{Milgrom_2024}.

\subsection{The RAR and MOND fits to galaxy rotation curves}

\begin{figure}[t]
\centering
\includegraphics[width=0.7\textwidth]{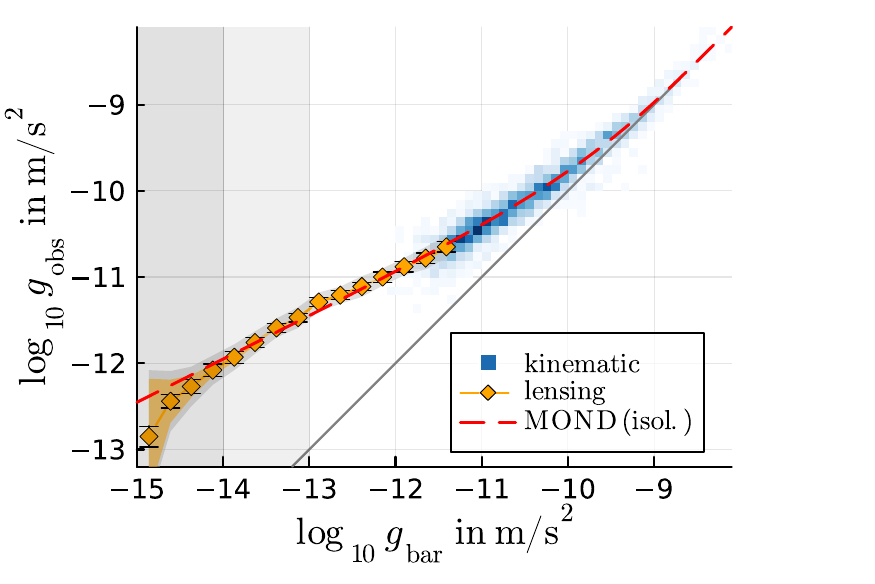}
\vspace{-0.3cm}
\caption{The kinematic RAR from \citet{LelliRAR} (in blue) and the lensing RAR from \citet{tobias} (in orange) together with the isolated MOND prediction of Eq.~\eqref{eqnu} and Eq.~\eqref{interpolating} (with $\delta=1$) in red. The grey line is the one-to-one line, for which the acceleration generated by baryons in Newtonian gravity and the observed one would be equal.}
\label{fig3}
\end{figure}

\begin{figure}[t]
\centering
\includegraphics[width=0.7\textwidth]{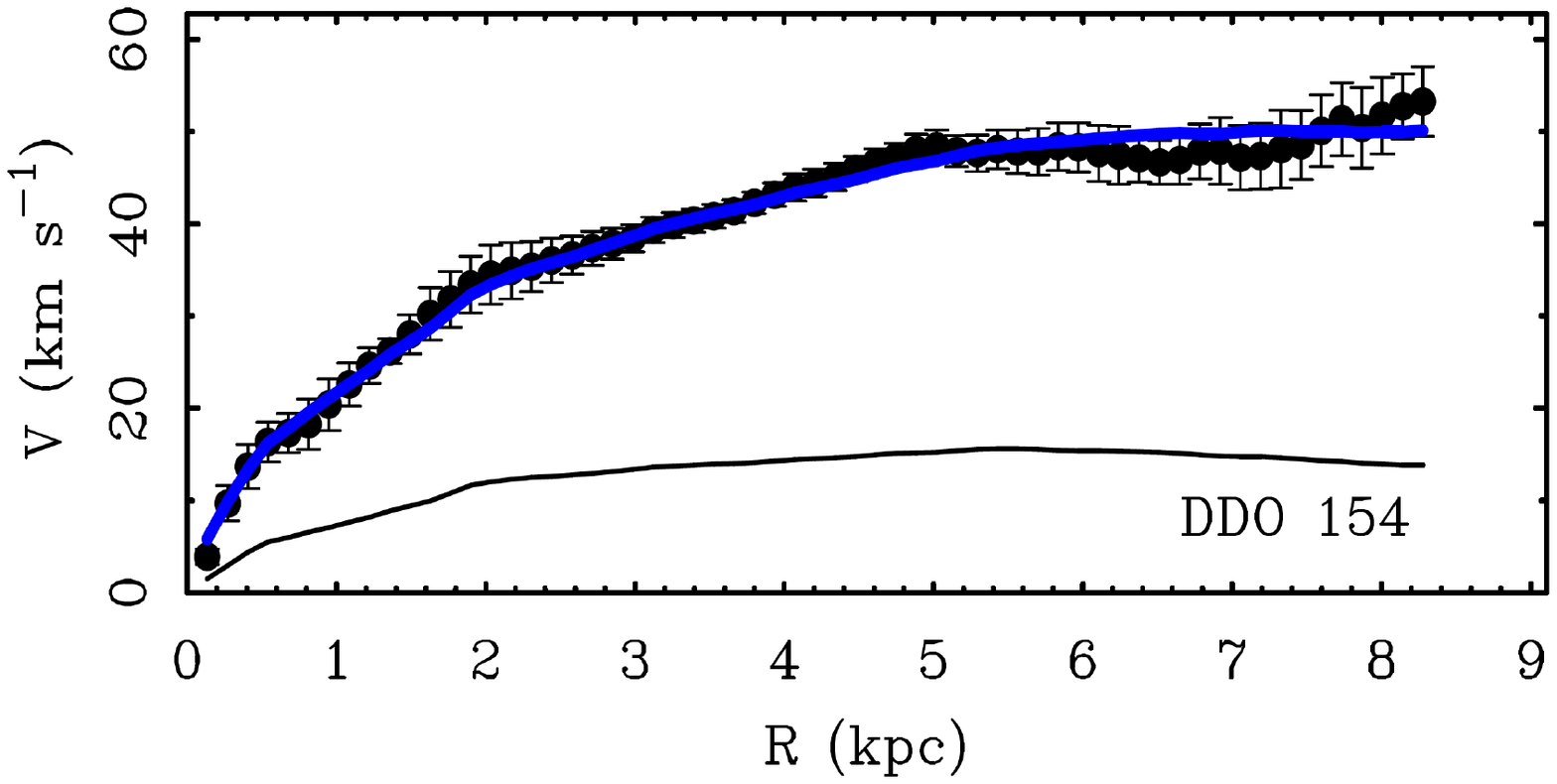}
\vspace{-1.4cm}
\caption{The observed rotation curve (points and error bars), Newtonian rotation curve generated by baryons (black), and MOND-predicted rotation curve (blue) for the LSB gas-dominated galaxy DDO~154 \citep{Famaey_2012}. This MOND-predicted rotation curve does not depend much on the stellar mass-to-light ratio nor on the interpolating function.}
\label{fig4}
\end{figure}

The BTFR and the CSDR can, in fact, both be deduced from a single, apparently more fundamental, local scaling between the measured gravitational acceleration at any point in the disk of rotationally-supported galaxies and the one generated by the observed baryon distribution, which --- amazingly --- cannot be tightened by including any other available galaxy property \citep{stis}: This relation is known as the {\it Radial Acceleration Relation} \citep[RAR,][]{McGaughRAR,LelliRAR}, and is in fact {\it exactly} equivalent to the MOND relation of Eq.~\eqref{eqnu}. At the phenomenological level, it can be considered as the local version of the BTFR: it implies the BTFR, but the BTFR does not imply it. As a matter of fact, it means that the inner parts of LSB galaxies that are in the low acceleration regime sit on  the same relation as the outer parts (of both LSB and HSB galaxies) when plotting the total radial gravitational acceleration vs. that generated by baryons in Newtonian gravity. This relation is actually successful beyond the realm of rotationally supported disk galaxies, extending to ellipticals \citep{LelliRAR} and even very low accelerations from weak lensing measurements around galaxies \citep{tobias}, as illustrated on Fig.~\ref{fig3}, a result which, in MOND, implies having a relativistic theory (see Sect.~4) in which lensing and dynamics are governed by the same potential in the weak-field static limit. 

This is in line with the central prediction of \citet{mil83b}, namely that {\it velocity curves calculated with the modified dynamics on the basis of the observed mass in galaxies should agree with the observed curves}. In practice, the rotation curve fits in MOND involve small uncertainties on the galaxies' parameters such as inclination $i$ (zero for face-on disks), distance $D$, and mass-to-light ratios $\Upsilon$ of the different stellar components (and even gas component), which can all be fit in a Bayesian fashion. Concretely, starting from rotation curves computed for flattened disks from the Newtonian Poisson equation for mass-to-light ratios of 1, one can adjust the mass-to-light ratios by renormalizing the Newtonian acceleration as
    \begin{align}
        a_{\rm N}\left(R \right)= \MLg \frac{\Vg^2}{R} + \MLd \frac{\Vd^2}{R} + \MLb \frac{\Vb^2}{R} \, .
   \end{align}
    Note that there is not much room to play around with the value of $\MLg$ which is much better known than its stellar counterparts. This first step is independent of any distance scale because $V^2$ and $R$ both scale as $D$. Then, using Eq.~\eqref{eqnu} to get $a(R)$ in MOND, one can adjust the original distance by a factor $\alpha = D_{\rm fit}/D$ (where $D$ is the measured distance, with priors coming from the observational uncertainties) and fit an inclination $i_{\rm fit}$ (also using priors from observational uncertainties on $i$) through
    \begin{align}
        V(R \alpha) {\sin i} = \sin i_{\rm fit} \sqrt{a(R) R \alpha }.
    \end{align}
One can additionally take into account the EFE through various approximations given in, e.g., Table~1 of \citet{ChaeMil}. Using parametric forms of the interpolating function such as 
\begin{subequations}\label{eq:if_fams}
	\begin{eqnarray}
    \nu_n(x)&=&\left[\frac{1+\left(1+4x^{-n}\right)^{1/2}}{2}\right]^{1/n}\, ,  \label{eq:nun}\\
    \nu_\delta(x)&=&\left(1-e^{-x^{\delta/2}}\right)^{-1/\delta}, \label{eq:IF_delta}
	\end{eqnarray}
 \label{interpolating}
\end{subequations} \citet{Desmond24} showed that the best fits to galaxy rotation curves were systematically close to $n \simeq 1$ or $\delta \simeq 1$, which corresponds to the traditional \citep{LelliRAR} form of the RAR. This conclusion remains broadly true when using Eq.~\eqref{aqualRC}, although the fits are slightly less good. It is therefore interesting to note a slight preference for the straight algebraic relation of Eq.~\eqref{eqmu} (and exact for circular orbits only in modified inertia) over the modified gravity correction, although not with high significance. The only exception to the $\delta \simeq 1$ best-fit form of the interpolating function is when removing galaxies with bulges and allowing the stellar mass-to-light ratios to float freely, yielding higher $\MLd$ and a much sharper best-fit transition with $\delta\simeq 2.5$, but incompatible with the rotation curves of bulgey galaxies. Moreover, allowing a galaxy-by-galaxy EFE is strongly disfavoured by the Bayesian Information Criterion relative to the no-EFE fit due to the addition of too many free parameters, and allowing a single global EFE strength is mildly disfavoured. There is therefore no strong evidence for a signature of the EFE in galaxy rotation curves. The EFE however remains {\it a priori} necessary to explain, for instance, the escape speed curve of the Milky Way \citep[e.g.,][]{Oria}. It is worth pointing out that the rotation curves of LSB gas-dominated galaxies provide, by far, the best test of the MOND phenomenology, by eliminating both the uncertainties relating to stellar mass-to-light ratios and that of the exact shape of the interpolating function: those galaxies (see Fig.~\ref{fig4} for an example) indeed follow the MOND prediction with impressive accuracy \citep[][]{Sanders2019}. Moreover, a last prediction of MOND (in its modified gravity realization) concerning galaxy rotation curves is that, in polar ring galaxies where two velocity curves can be measured in two perpendicular planes, velocities are predicted to be higher in the extended polar ring than in the more compact host galaxy, in accordance with observations \citep{polar}. The reason is that the host appears more compact and spherical to the orbits within the ring than the extended ring appears to orbits within the host at the same radii. This behaviour of the velocity curves is expected in Newtonian gravity without dark matter, and is boosted in MOND, but should have been mostly washed out in the presence of a common spherical dark matter halo.

\subsection{Spheroidal pressure-supported systems}

Most of the observational success of MOND hereabove concern late-type galaxies, but this is in part because those have much more precise tracers of their gravitational field, such as rotation curves. The case of early-type galaxies is less clear, but isolated ones do appear to follow the same RAR as disk galaxies, as evidenced by the kinematics of early-type galaxies that have inner rotating stellar components and outer HI rings, and from the hydrostatic equilibrium of those that have relaxed X-ray emitting haloes \citep{LelliRAR}. Moreover the weak-lensing RAR of \citet{tobias} displayed on Fig.~\ref{fig3} is identical for early-type and late-type galaxies (with a careful treatment of stellar mass-to-light ratios). At slightly larger scales (and lower accelerations), intermediate-richness galaxy groups also show remarkable agreement with MOND \citep{Milgromgroups}. At smaller scales, concerning, e.g., dwarf spheroidal galaxies, MOND has had success in predicting the velocity dispersions of the satellites of M31 \citep{McMil}, but in dwarf spheroidal satellites of the Milky Way, especially in the faintest ones, the situation is less clear \citep[e.g.,][]{lugdsph}. In general, those ultra-faint galaxies appear too dark matter dominated to be in accordance with MOND, if they truly are at equilibrium. The reverse appears to be the case in some globular clusters \citep{ibata} not displaying enough of a gravitational boost \citep[but see][]{sanders2419}. The observations of ultra-diffuse galaxies lacking dark matter in Newtonian gravity, on the other hand, present only a very mild tension with MOND once taking into account the EFE \citep{Famaey_2018,Muller}.

While these small tensions in spheroidal systems are actually not entirely unequivocal because of modelling uncertainties, the apparent failure of MOND is most evident in galaxy clusters: in the central parts of clusters, gravitational accelerations typically exceed the $a_0$ acceleration scale by an order of magnitude. This would mean no gravitational boost in MOND, whilst dark matter {\it is} needed there in Newtonian gravity, to explain the temperature profile of X-ray emitting gas, the kinematics of galaxies and strong lensing signatures from the inner parts of clusters. This is known as the residual missing mass problem of galaxy clusters in MOND \citep[e.g.,][]{angusbuote,kelleher}. The total residual missing mass in clusters is roughly equivalent to the observed baryonic mass (hot gas and stars) and has actually been shown from cluster lensing \citep{Famaey25} to have a mass density profile $\rho_{\rm res}$ following the hot gas mass profile $\rho_\text{gas}$ with
\begin{align}
\label{eq:propto}
\rho_{\rm res}(r) = \eta \, \rho_\text{gas}(r) {\rm exp}(-r/r_{\rm cut}),
\end{align}
where $\eta$ is the ratio of residual missing mass to gas mass in the inner parts and $r_{\rm cut}$ the cut-off radius. The most straightforward interpretation  of this discrepancy in the modified gravity context would be that MOND {\it predicts} additional baryonic mass in the center of galaxy clusters, e.g. in the form of numerous cold, dense ${\rm H}_2$ gas clouds. This is in line with the observation that the missing mass follows the hot gas distribution in the central parts. Another hypothesis stems from the fact that MOND is not {\it per se} incompatible with the existence of cosmologically relevant hot non-baryonic dark matter (e.g., sterile neutrinos of a few eV). The density of such hot dark matter would be relevant only on galaxy cluster scales and above, but the results of simulations actually greatly reduce the appeal of the idea \citep[][see Sect.~3.6]{angussim,wittenburg}. Finally, the apparent failure of MOND in galaxy clusters can be seen as a reminder that MOND is only a paradigm in need of embedding in a more fundamental theory, that could involve other scales than acceleration (e.g., Sect~2.6), as well as multiple new fields. It can also be interesting to visualize the residual missing mass problem of MOND by looking at the RAR of galaxy clusters from detected baryons: the residual missing mass appears as a hump in the RAR at high accelerations, followed by a RAR with a larger normalization than the one of Fig.~\ref{fig3}. Another, less clear, potential tension for MOND in clusters appears in the shape of the RAR at low accelerations, apparently falling below the MOND expectation: this tension becomes milder when taking into account the EFE \citep{kelleher},  whilst \citet{Durakovic} showed that some relativistic versions of MOND that induce oscillations of the gravitational field could automatically provide such a sharp drop at large radii. But perhaps the most iconic challenge for MOND in galaxy clusters is the one posed by the archetypal Bullet Cluster 1E0657-558. In this cluster, a subcluster of X-ray emitting plasma peak is separated by $\sim 300$~kpc on the sky from the main emitting plasma peak, and displays a prominent bow shock indicating that it collided with the main plasma peak about 100~Myr prior to the current configuration. The galaxy concentrations, representing only 15\% of the baryonic mass of the cluster, are on the other hand separated by 720~kpc on the sky, i.e., they are offset from the plasma peaks towards the outskirts of the system, indicating that the galaxies were not slowed down by the shock, as expected. The lensing convergence map is centered on the galaxy concentrations, which is consistent with collisionless dark matter following the trajectory of the galaxies. In other words, the lensing peaks are offset from the main peaks of baryonic surface density. In MOND, it {\it is} actually expected, in general, that the phantom mass density will not align with the baryonic density. After all, the phantom density around a disk galaxy is spherical at large radii, and most of the phantom mass is where baryons are not. In specific three-peak configurations, it {\it is} possible, for a given lensing potential, that the corresponding two outer baryonic surface density peaks in MOND would be more inwards than their phantom counterpart. However, a relativistic theory reducing to AQUAL (or QUMOND) in the quasi-static limit is not sufficient to explain the Bullet Cluster lensing. The residual missing mass density is typically the same as in relaxed clusters, and is indeed centered on the galaxy concentrations \citep{Angus2007}. This would be in line with the residual missing mass being made of numerous cold, dense gas clouds (which would behave as galaxies in the encounter) or any form of hot non-baryonic dark matter. To explain away those leaning peaks with phantom density alone, the only way out would probably be to rely on the fact that the subclusters are moving at about 1\% of the speed of light and that the situation is not, strictly speaking, that of the quasi-static limit. In a multifield theory, it is possible that baryons and field perturbations would not propagate at the same speed. Reproducing the {\it details} of the lensing map of the Bullet Cluster, with the right proportion of residual missing mass, could still prove very challenging, though.

\subsection{Small scales}
The most obvious place to probe a modification of gravity in a direct/experimental way is of course the Solar System. The first MOND correction to appear in the inner Solar system should manifest itself as a (tiny) anomalous gravitational field in addition to the Newtonian one. The anomalous acceleration is given by $a(1-\mu)$ or $a_N(\nu-1)$, which can be constrained from the motion of inner planets (perihelion precession and non-variation of Kepler's constant). If taking $n=1$ in the first family of interpolating functions in Eq.~\eqref{interpolating}, one has that $x(\nu(x)-1) \to 1$ for $x \to \infty$, meaning that the anomaly is a constant acceleration of amplitude $a_0$. This is ruled out since the anomalous acceleration is constrained to be smaller than $\sim a_0/10$ on the orbits of Mars or Jupiter. The $\delta=1$ interpolating function does not suffer from the same problem. However, in AQUAL and QUMOND, another anomaly is generated by the EFE from the Milky Way (of the order of $a_{\rm ext} \sim 1.8 a_0$). Even when the internal field largely dominates over the external one, the internal phantom density flattens along the external field direction, and modifies the Newtonian potential of the Sun by a quadrupole anomaly $\delta \Phi(\vec{x}) = -Q_2 x^i x^j\left(\hat{e}_i \hat{e}_j - \delta_{ij}/3 \right)/2$, where ${\hat{e}}= {\vec{g}}_{\rm ext}/g_{\rm ext}$ is a unit vector pointing towards the Galactic center, $\delta_{ij}$ the Kronecker delta, $\vec{x}$ the position within the Solar System with respect to the Sun, and, in QUMOND \citep[see][and references therein]{Desmond24},
\begin{equation}\label{eq:q}
    Q_2 = -\frac{9a_0^{3/2}}{4\sqrt{G{\rm M}_\odot}}\int_0^\infty \mathrm{d} v \int_{-1}^1 d\xi \left(\nu -1\right)\Big[e_{\rm N}\left(3\xi -5\xi^3\right)+v^2\left(1-3\xi^2\right)\Big] \, ,
\end{equation}
where $\nu=\nu\left[\sqrt{e_{\rm N}^2+v^4+2e_{\rm N}v^2\xi}\right]$ and $e_{\rm N}$ is the solution of $e_{\rm N}\nu\left(e_{\rm N}\right)=a_{\rm ext}/a_0$. The value of $Q_2$ is typically predicted to be slightly larger in AQUAL than in QUMOND. Nine years of Cassini range and Doppler tracking data have observationally constrained it to be $Q_2 = \left(3\pm 3\right) \times 10^{-27} \, \mathrm{s}^{-2}$, at one sigma \citep{Hees2014}. This also rules out the $\delta=1$ transition of Eq.~\eqref{interpolating}, that best describes galaxy rotation curves, but not a sharper $\delta \simeq 2.5$ transition which predicts a very low $Q_2$. Most likely, these constraints \citep[see also][for similar constraints from the distribution of binding energies of long-period and Oort-cloud comets detectable from Earth]{comets} imply that modified gravity MOND needs a new scale in addition to acceleration (see Sect.~2.6) to pass Solar System constraints, or that MOND rather results from a more radical modification of inertia (see Sect.~2.7). Because the $Q_2$ prediction of AQUAL/QUMOND mainly probes the interpolating function close to $e_{\rm N}=a_\mathrm{N,ext}/a_0$, any transition that passes this test in QUMOND or AQUAL also predicts zero deviation from Newtonian dynamics in wide binaries of the Solar neighbourhood, whose deviation from Newtonian dynamics would be primarily governed by an effective renormalization of the gravitational constant from the Galactic EFE. While the jury is still out on whether wide binaries in the Solar neighbourhood rule out the AQUAL/QUMOND behaviour expected for a $n=1$ or $\delta=1$ interpolating function \citep{Banik24,Chae24,Hernandez}, any confirmed detection of such a behaviour would be inconsistent with Solar System constraints in that very same AQUAL/QUMOND framework (but not necessarily with their generalizations outlined in Sect.~2.6).

\subsection{MOND simulations}

As outlined in Sect.~2.4, N-body/hydrodynamical codes implementing the modified gravity MOND field equations, such as \textsc{PoR} \citep{Lughausen_2015, Nagesh_2021} have been developed \citep[both in QUMOND and AQUAL, see also][]{Candlish}. Such codes have allowed to simulate, for instance, the dynamics of isolated disks, showing for instance that bars are easily formed and tend to be fast in MONDian stellar disks, whilst they can be either fast or slow when including gas \citep{Tiret,Nagesh_sims}, that MONDian gas-rich clumpy galaxies in the early universe do not form bulges \citep{Combes}, or that the morphology of tidal tails \citep{Renaud} or shells \citep{Bilek} triggered by galaxy interactions can be reproduced in MOND. Perhaps the most interesting result coming out of such simulations has been related to systems sensible to the EFE in the $|\vec{a}_{\mathrm{ext}}| \sim |\nabla \Phi_{\mathrm{int}}|$ regime, namely the lopsidedness of the internal gravitational field of globular clusters or open clusters orbiting around the Milky Way, and the resulting {\it asymmetry} of their tidal streams, in accordance with observations \citep{Thomas18,Kroupa22}. On the other hand, simulations of structure formation in MOND have been limited to a very narrow context because MOND, {\it as it is}, is mostly mute on cosmology. The only context in which such simulations have been carried out is when combining MOND with hot non-baryonic dark matter. This hypothesis has the advantage of allowing to devise simulations of structure formation in a MOND context independently of a particular relativistic theory, by assuming that the dynamics is strictly in the GR regime at $z>200$ (which is a working hypothesis but does not have to be the case in general). Such simulations systematically overproduce high mass structures at all redshifts \citep{angussim,wittenburg}. Although it has been qualitatively argued that this mismatch might be reduced by considering that we live in an underdensity, it still reduces the appeal of such models. It might be much more fruitful to explore the consequences of relativistic realisations of MOND in cosmology, which will be the topic of the next section.

\section{Relativistic theories and cosmology}

\subsection{A relativistically invariant theory}
As outlined in Sect.~3.5, the Solar System provides strong constraints on departures from GR. If there is an additional contribution (a `fifth force') it must be screened, i.e., be made undetectable in currently probed regimes. Observations of gravitational waves also constrain departures from GR, in particular the departure of the speed of tensor waves from the speed of light, which with the observation of electromagnetic counterparts, provides forbidding constraints.
It is therefore natural to look for a theory that has a GR-like limit. It is also theoretically well-motivated as GR is the unique four-dimensional local massless spin-2 field theory coupling to the energy-momentum tensor with equations of motion of second order. Other theories necessarily incorporate extra degrees of freedom or break locality or add extra dimensions. While a non-relativistic theory of MOND was straightforward to formulate, since gravity itself is described by a scalar field, it is not straightforward to generalise starting with GR. The potential itself is only, in the non-relativistic limit of slow motions the diagonal piece of the metric $g_{\mu \nu} = \eta_{\mu \nu} + h_{\mu \nu}$ where $h_{00} = h_{ii} = 2 \Phi$. MOND is an acceleration-based modification of gravity, but the connection, which is the object `playing the role of acceleration' in the weak-field limit of GR, is not a tensor. In the search for a fundamental theory of MOND it is perhaps prudent to keep in mind that GR itself was not a simple generalisation of Laplacians, $\nabla^2 $ to d'Alembertians $\nabla^2 \to g^{\mu\nu}\nabla_\mu \nabla_{\nu}$, nor only partial derivatives to covariant derivatives but required the development of the more sophisticated mathematics of differential geometry.

The action of GR is the sum of the matter action and the Einstein--Hilbert (gravitational) action, the Lagrangian density being :
\begin{align}
\mathcal{L} = \mathcal{L}_\mathrm{matter} + \frac{c^3}{16 \pi G} (\mathcal{L}_\mathrm{EH} - 2 \Lambda \sqrt{-g}),
\label{lagGR}
\end{align}
with $\Lambda$ the cosmological constant, and 
\begin{align}
\mathcal{L}_\mathrm{EH} = R \sqrt{-g}\, , 
\end{align}
where $g$ denotes the determinant of the metric tensor matrix $g_{\mu\nu}$ with $(-,+,+,+)$ signature, and $R=R_{\mu\nu}g^{\mu\nu}$ its scalar curvature (the Ricci scalar), $R_{\mu\nu}$ being the Ricci tensor (involving second derivatives of $g_{\mu\nu}$). The most obvious way to extend it is to add new fields on top of the metric, each adding their own Lagrangian density on top of $\mathcal{L}_\mathrm{EH}$ within the parenthesis of the second term of Eq.~\eqref{lagGR}. Such additional fields will typically break the Strong Equivalence Principle, which is desirable in MOND.

\subsection{Scalar-Tensor theories}

\subsubsection{Relativistic AQUAL}

The first relativistically invariant proposal was already proposed in the appendix of the original AQUAL paper \citep{BM84}, by considering a scalar-tensor theory with a non-canonical kinetic term in the Lagrangian density of the scalar field $\phi$
\begin{align}
\mathcal{L}_{\phi} \propto \sqrt{-g}\, f(X),
\end{align}
where $X \propto \nabla^\nu \phi \nabla_\nu \phi$ and the function $f(X) \propto X^{3/2}$ at small $X$, the hallmark of a MOND Lagrangian. Matter couples to ${\rm exp}(-2\phi) g_{\mu\nu}$, i.e. the `physical' metric of the matter frame is a conformal transformation of the Einstein metric, affecting all components in the same way. In such a framework, causality can be preserved under appropriate conditions on the function $f$, but gravitational lensing is insensitive to the conformal rescaling of the metric. In other words, while galaxy rotation curves would have an AQUAL-like behaviour, such a theory would not reproduce the lensing results displayed in Fig.~\ref{fig3}. Note that the potential governing weak-field dynamics ends up being the sum of the Newtonian potential and a second one obeying an AQUAL-like equation. To recover MOND dynamics, the function $f$ must be chosen in such a way that its corresponding interpolating function $\hat{\mu}$ is related to the AQUAL interpolating function $\mu$ through $\hat{\mu}(y) = (x-y)/y \; {\rm where} \; y \equiv x[1-\mu(x)]$, which has the same deep-MOND asymptotic behaviour as $\mu$. Whilst such theories would {\it a priori} struggle to pass the Solar System constraints of Sect.~3.5, it is possible to construct a theory that introduces an additional length-scale (in the spirit of Sect.~2.6) by adding a Galileon-type term to the scalar field Lagrangian, which allows to completely screen MOND effects on small scales \citep{Babichev}.

\subsubsection{Phase coupling gravitation}

While adding a real scalar field needs the inclusion of a non-canonical kinetic term to reproduce MOND in the weak-field limit, this is not the case for a complex one, $\xi = q \,{\rm exp}({\rm i} \, \phi)$. Phase coupling gravitation \citep[PCG,][]{PCG} considers that only the phase $\phi$ couples to matter jointly with the metric\footnote{A related idea \citep{superfluid} has been to propose a theory in which dark matter Bose-Einstein condenses into a superfluid phase in the central regions of halos. Such a superfluid phase is described by the theory of spontaneously broken global $U(1)$ symmetry, with an order parameter field $\xi = |\xi| \, {\rm exp}({\rm i}\phi)$ where the phonon field $\phi$ is the relevant degree of freedom, and couples to baryons, as in PCG. The effective field theory is then essentially described by a scalar field Lagrangian with a non-canonical kinetic term, for which an AQUAL-like action with $3/2$ power index can be assumed or derived from a sextic potential of $|\xi|$. Outside of the superfluid core, dark matter would behave as CDM. Such a theory leads to a rich phenomenology described in \citet{superfluidpheno}, recovering by design most of the MOND phenomenology in galaxies, but potentially struggling to reproduce lensing observations.}, through ${\rm exp}(-2\phi) g_{\mu\nu}$, and that the Lagrangian density possesses a standard self-interaction potential $V(q)$, namely
\begin{align}
\mathcal{L}_{\xi} \propto \sqrt{-g}\, \left( \frac{q^2}{2} \, \nabla^\nu \phi \nabla_\nu \phi + \frac{1}{2}\, \nabla^\nu q \nabla_\nu q + V(q) \right).
\end{align}
Variation with respect to $q$ then yields $\nabla^\nu \phi \nabla_\nu \phi=-V'(q)/q$, and variation with respect to $\phi$ yields and AQUAL-like equation with $q^2$ playing the role of the interpolating function. One typically recovers MOND in the static weak-field limit if the potential is sextic, $V(q) \propto -q^6$, which effectively gives back the hallmark non-canonical kinetic term $(\nabla^\nu \phi \nabla_\nu \phi)^{3/2}$ in the Lagrangian.

\subsection{Non-dynamical vector fields}

None of the above scalar-tensor theories can, however, reproduce the lensing observations of Fig.~\ref{fig3}. There is a simple cure to this problem, namely to add a non-dynamical timelike vector field $A_\mu$ with unit-norm, in order to enforce a non-conformal relation between the Einstein and physical metrics, such that matter couples to ${\rm exp}(-2\phi) g_{\mu\nu} - 2\sinh(2\phi) A_\mu A_\nu$. Such a vector field can be added on top of the relativistic AQUAL \citep{Sanders97} as well as on top of the PCG formalism \citep[Bi-scalar-tensor-vector theory, BSTV,][]{Sanders05}. However, it is desirable to let the fields of the theory all be dynamical (notably because theories then tend to be safer regarding instabilities).

\subsection{Tensor-Vector-Scalar theory}

The idea of the Tensor-Vector-Scalar \citep[TeVeS,][]{teves} theory is to still use a vector field in order to keep the former non-conformal relation between the Einstein and physical metrics, but to replace the non-dynamical vector field by a dynamical one, $A_\mu$, with a Lagrangian similar to that of the electromagnetic 4-potential vector field, but without the coupling term to the 4-current, and with a Lagrange multiplier forcing the unit norm. The Lagrangian density of the scalar field keeps its relativistic AQUAL form, with a power 3/2 non-canonical kinetic term, or can alternatively make use of a second non-dynamical (i.e., with no kinetic term) scalar field $\mu$ playing the exact same role as $q^2$ in PCG hereabove. When rewriting the Lagrangian of TeVeS entirely in the matter frame, one can recast it as that of a \emph{non-unit norm} vector field, $B_\mu$, such that $B^{2} \equiv g^{\mu \nu}B_{\mu}B_{\nu} = -{\rm exp}(-2\phi)$. This Lagrangian density then depends {\it only} on the non-unit norm vector field, the non-dynamical field $\mu$, and a dimensionless parameter $K_B$:
\begin{align}
\mathcal{L}_{B} \propto \sqrt{-g}
          \left[K^{\alpha \beta \mu \nu}\nabla_{\alpha}B_{\mu}\nabla_{\beta}B_{\nu}
          +\frac{V(\mu)}{B^{2}}\right],
\end{align}
where 
\begin{align}
K^{\alpha \beta \mu \nu} = d_{1}g^{\alpha \beta}g^{\mu \nu} + d_{2}g^{\alpha \mu}g^{\beta \nu}+ d_{3}g^{\alpha \nu}g^{\beta \mu} 
+d_{4}B^{\alpha}B^{\beta}g^{\mu \nu}+d_{5}
g^{\alpha \nu}B^{\beta}B^{\mu}+d_{6}g^{\alpha \beta}B^{\mu}B^{\nu}
+d_{7}g^{\alpha \mu}B^{\beta}B^{\nu}+
d_{8}B^{\alpha}B^{\beta}B^{\mu}B^{\nu},
\label{di}
\end{align}
and the $d_i$ coefficients depend on $B^2$, on the scalar $\mu$, and on the parameter $K_B$ \citep{Zlosnik2006}. A key aspect of TeVeS is that the speed of the tensor mode gravitational waves is generically different than the speed of light, and it has thus been ruled out after GW170817, which provided a limit on the difference between the speed of light and that of gravitational waves. However, note that freely choosing the dependence of the $d_i$ coefficients in Eq.~\eqref{di} straightforwardly generalizes the original TeVeS, offering the hope to possibly evade such a constraint.

\subsection{Aether-Scalar-Tensor theory and cosmology}

The Aether-scalar-tensor theory \citep[AeST,][]{aest} has the same field content as TeVeS, namely the metric $g_{\mu \nu}$, a unit time-like vector field $A_{\mu}$ and a scalar field $\phi$. It can be seen as a generalisation of TeVeS that passes the speed of gravitational waves speed constraint, by appropriately choosing the coefficients (and differently to their TeVeS values) in the equivalent of Eq.~\eqref{di} for the unit-norm vector field $A_{\mu}$. To write down the AeST Lagrangian in its most `explicit' form, one can make use of the scalar gradients projected onto $A^{\mu}$,  as well as perpendicular to it, $\mathcal{Q} \equiv A^{\mu} \nabla_{\mu} \phi$ and $\mathcal{Y} \equiv (g^{\mu \nu} + A^{\mu} A^{\nu})\nabla_{\mu}\phi \nabla_{\nu}\phi$, respectively, and the vector gradient projected onto the vector itself $J_{\mu} \equiv A^{\nu} \nabla_{\nu} A^{\mu}$. These kinetic terms are lumped together in the TeVeS Lagrangian hereabove, but can be explicitly separated from each other within the AeST Lagrangian for the vector and scalar fields, which reads
\begin{align}
\mathcal{L}_B \propto \sqrt{-g} \left( - \frac{K_B}{2} F^{\mu \nu} F_{\mu \nu} + \left(2-K_B\right)\left( J^{\mu} \nabla_\mu \phi - \mathcal{Y} \right) - \mathcal{F}\left(\mathcal{Y},\mathcal{Q}\right) - \lambda \left(A^{\mu} A_{\mu} + 1\right)\right),
\end{align}
where $F_{\mu\nu} = \nabla_{\mu} A_{\nu} - \nabla_{\nu} A_{\mu}$ is the field strength of $A_{\mu}$, and $\lambda$ is a Lagrange multiplier enforcing the unit-norm constraint of $A_{\mu}$. In addition to the bare gravitational constant and $K_B$ the theory will have additional parameters entering the definition of the free function $\mathcal{F}$. To further simplify the expression of the Lagrangian, one can consider the separable case $\mathcal{F}(\mathcal{Y},\mathcal{Q}) \equiv \mathcal{J}( \mathcal{Y})+ \mathcal{K}(\mathcal{Q})$. As in the non-relativistic case of Sect.~2.1, the MOND behaviour for the quasi-static limit is incorporated through $\mathcal{J}(\mathcal{Y} ) \propto \mathcal{Y}^{3/2}/a_0$ for $\sqrt{\mathcal{Y} }\ll a_0$. When $\mathcal{K}(\mathcal{Q})$ has a minimum at a non-zero $\mathcal{Q}_0$, such as for the quadratic function $ \mathcal{K}(\mathcal{Q}) = \mathcal{K}_2 (\mathcal{Q} - \mathcal{Q}_0 )^2$, the scalar field displays a perfect fluid behaviour for displacements of $\mathcal{Q}\sim \dot{\phi}$ away from the non-zero value $\mathcal{Q}_0$. 

\begin{figure}[h]
\centering
\includegraphics[width=0.48\textwidth]{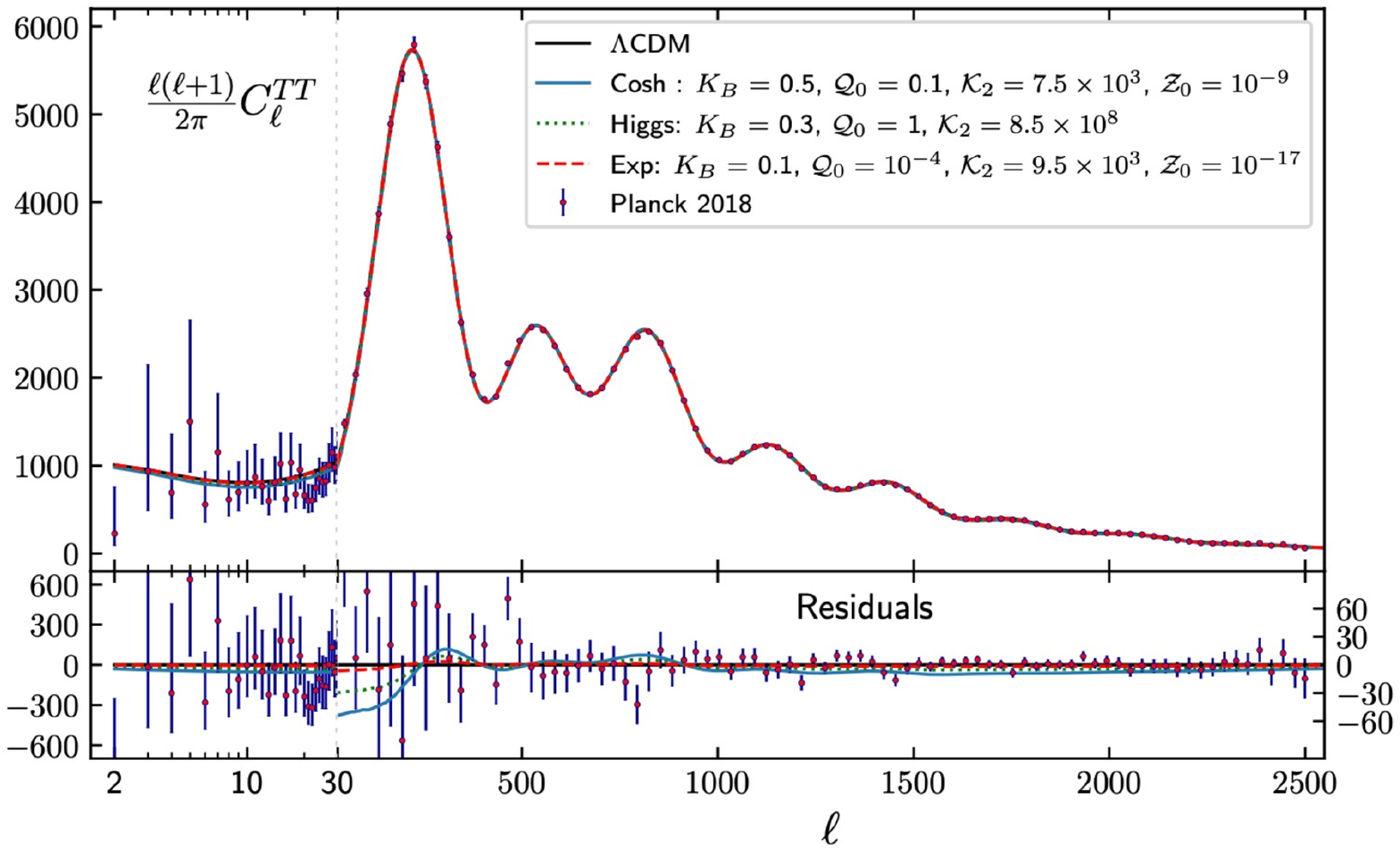}
\includegraphics[width=0.48\textwidth]{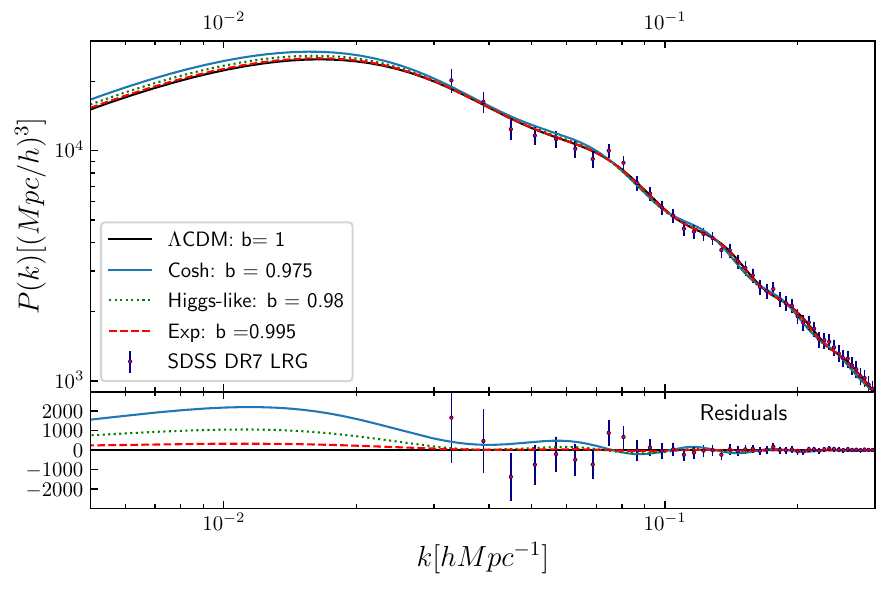}
\vspace{-0.3cm}
\caption{AeST can reproduce the angular power spectrum of the CMB (left panel) and linear matter power spectrum (right panel), with various choices of the $\mathcal{K}(\mathcal{Q})$ function \citep{aest}. }
\label{fig5}
\end{figure}

The cosmological evolution in AeST is to drive $\mathcal{Q}$ towards $\mathcal{Q}_0$, where $\dot{\phi} = \mathcal{Q}_0$, or $\phi(t) = \mathcal{Q}_0 t $. As this is the true vacuum, perturbations of the scalar field $\varphi(t,x)$ are expanded around it, $\phi(t,x) =\mathcal{Q}_0 t + \varphi(t,x)$.
The cosmological perturbations lead to a behaviour close to that of a perfect fluid that does not exactly reduce to the behaviour of particle dark matter: it mimics $\Lambda$CDM to linear order, but there are corrections, which lead to a pressure contrast, proportional to the speed of sound, which itself depends on the choice of the function $\mathcal{K}(\mathcal{Q})$. With AeST, MOND is not silent on cosmology anymore. The resulting angular power spectrum of the Cosmic Microwave Background (CMB) and matter power spectrum, for different choices of the $\mathcal{K}(\mathcal{Q})$ function, are displayed on Fig.~\ref{fig5}.

In the static weak-field limit of AeST, the MOND behaviour is explicitly seen to arise by making the metric ansatz $g_{\mu \nu}\mathrm{d}x^\mu \mathrm{d}x^\nu = - (1+\Psi) \mathrm{d}t^2 + (1+\Phi) \gamma_{ij} \mathrm{d}\vec{x}^2$, the vector decomposition $A_i = \nabla_i \alpha + \alpha_{\perp} $, considering a perfect fluid source, and expanding the action to second order in the fields. Taking variational derivatives then gives, $\Psi = \Phi$, so lensing is as in GR, and $\nabla^2 (\Phi- \chi ) + m^2 \Phi  = 4\pi G_{\mathrm{N} } \rho $, where $m^2 \equiv 2 \mathcal{K}_2 \mathcal{Q}_0^2/(2-K_B)$, $G_\mathrm{N} \equiv G/(2-K_B)$, and $\chi \equiv \varphi + \mathcal{Q}_0 \alpha$. Moreover, $\nabla^2 \Phi = \nabla \cdot (\beta(|\nabla \chi|)  \nabla \chi )$ where $\beta \equiv 1+ \mathcal{J}'(\mathcal{Y})$. When $m^2 \Phi$ can be neglected, $\Phi- \chi$ therefore plays the role of the Newtonian potential, and $\nabla \cdot ((\beta(|\nabla \chi|)-1)  \nabla \chi ) \simeq  4\pi G_{\mathrm{N} } \rho$, as it should in order to recover a MOND behaviour. The additional $m^2 \Phi$ term however typically becomes important at very large distances, and introduces oscillatory features in the solutions, which could lead to new behaviour in the outskirts of galaxy clusters \citep{Durakovic}.

\subsection{Khronons: back to a Scalar-Tensor theory}

In the Khronon theory, rather than directly using a dynamical vector field in the Lagrangian as in TeVeS or AeST, one instead introduces a preferred foliation of spacetime, consisting of hypersurfaces of constant value of a scalar field $\tau$, the Khronon field. The vector field is then derived from the scalar field $\tau$, by constructing the vector $n_{\mu} = - \nabla_\mu \tau/\mathcal{Q}$ where $\mathcal{Q} = \sqrt{-g^{\mu \nu} \nabla_{\mu} \tau \nabla_{\nu} \tau }$. As $n_{\mu}$ is automatically normalised it is not necessary to introduce a Lagrange multiplier in the action.
The AeST-like vector $A_{\mu} = n^{\nu} \nabla_{\nu} n_{\mu} $ has the meaning of being the four-acceleration of the set of worldlines with velocity $u_{\mu}$. The norm-square of this acceleration is called $\mathcal{Y} \equiv A^{\mu}A_{\mu}$.

The original Khronon action
reads:
\begin{align}
\mathcal{L}_\tau \propto - \sqrt{-g} \, \mathcal{J}(\mathcal{Y}) .\end{align}
As in AeST, there is a direct correspondence between the function $\mathcal{J}$ and the MOND interpolating function $\mu$, the deep-MOND regime corresponding to $\mathcal{J}(\mathcal{Y})\to -\mathcal{Y} + 2 \mathcal{Y}^{3/2}/(3a_0)$. Also, when expanding in weak fields, $\Phi = \Psi$, so lensing is as in GR. There are two formulations of the theory: a $3+1$ formulation in `adapted coordinates' where $\tau$ is the time coordinate $t$, and a $4D$ formulation where it is a dynamical field. This model can be given dust-like behaviour in cosmology, as in AeST, by the addition of a kinetic term for the Khronon $\mathcal{K}(\mathcal{Q})$ to the action so that \citep{Blanchet_2024}
\begin{align}
\mathcal{L}_\tau \propto - \sqrt{-g} \, \left( \mathcal{J}(\mathcal{Y}) + \mathcal{K}(\mathcal{Q}) \right)
\end{align}
where $\mathcal{K}(\mathcal{Q})$ has a non-zero minimum. As in AeST, the choice of $\mathcal{K}(\mathcal{Q})$ determines the cosmological evolution of the dark fluid. The energy-density is given by $\rho = (\mathcal{Q} \mathcal{K}_{\mathcal{Q}}- \mathcal{K})/(8 \pi G)$ and pressure $p = \mathcal{K}/(8 \pi G)$ with adiabatic speed of sound $c^2_{\mathrm{ad}} = \mathcal{K}_{\mathcal{Q} } / (\mathcal{Q} \mathcal{K}_{\mathcal{Q}\mathcal{Q}} )$. Hence, even at linear order there are departures from $\Lambda$CDM cosmology, but they can remain at levels compatible with observations depending on the choice of $\mathcal{K}$. It is remarkable that a theory based on only two dynamical fields, the metric and the scalar Khronon field, can recover MOND at the scale of galaxies, for both lensing and dynamics, as well as a cosmology compatible with observations in the linear regime.

\subsection{Bi-metric theories}
It is also possible to recover MOND by adding a second rank-two tensor, i.e. a second metric $\hat{g}_{\mu \nu}$, instead of new scalar or vector degrees of freedom. Indeed, in GR it is the connection that plays the role of acceleration, but one can construct a tensor out of it only if there is more than one metric. In that case, the difference between the associated connections is indeed a tensor, and a scalar with dimensions of acceleration can be constructed out of it. Denoting the auxiliary metric $\hat{g}_{\mu \nu}$, to which some `twin matter' might couple, one can then, for instance, add to the GR Lagrangian the following
\begin{align}
{\cal L}_{\hat{g}} \propto \sqrt{-g} \, [{\hat{R}} - f(X)],
\label{bimonda}
\end{align}
where $\hat{R}$ is the Ricci scalar associated to the second metric, and $X \propto g^{\mu \nu} (C^{\alpha}_{\mu \beta} C^{\beta}_{\nu \alpha} - C^{\alpha}_{\mu \nu}C^{\beta}_{\beta \alpha})$, with  $C^{\alpha}_{\mu \nu} = \Gamma^{\alpha}_{\mu \nu} -  \hat{\Gamma}^{\alpha}_{\mu \nu}$ being the tensor playing the role of acceleration. The modification of gravity comes from the interaction between the spacetime on which matter lives and the auxiliary spacetime on which some `twin matter' might live. The above example is only one out of a vast class of bi-metric relativistic versions of MOND \citep[BIMOND,][]{Milgrom_bimond}.

\section{Outlook}

This chapter presented the MOND paradigm of modified dynamics and its large range of successes on galaxy scales. These phenomenological successes certainly do call for an explanation. The most mundane one, in the dark matter context, would be that this phenomenology will naturally emerge from a full-fledged understanding of the physics of baryons and its associated feedback within galaxies. However, it is not unlikely that this phenomenology might also teach us something fundamental that we are still missing about the nature of the dark sector itself, and possibly involve a modification of dynamics on galaxy scales, as prescribed by MOND. It has been shown how multi-field relativistic theories can in principle provide an effective framework within which such a modification is realized, whilst reproducing basic cosmological observables such as the angular power spectrum of the CMB. Much more radical ideas have however also been proposed. For instance, it is possible to construct pure metric-based modified gravity theories doing away with dark matter, but at the price of abandoning locality \citep{Deffayet} or covariance \citep[][]{noncov}. The latter idea can also be related to non-linear extensions of the Coincident General Relativity formulation \citep{Dambrosio}. Even more radical ideas arise once considering a modified inertia description of MOND, where the inertial term $m a$ would become $m a^2/a_0$ in the MOND regime for circular orbits. Relying on an adaptation of the Le Chatelier principle, \citet{Milgrom99} for instance proposed that, in de Sitter spacetime, the inertial term could actually become proportional to the difference between the Unruh temperature seen by an accelerated observer, $T_U \propto \sqrt{a^2 + a_0^2/4}$, and the Gibbons-Hawking one, $T_{\Lambda} \propto a_0/2$, where one has defined $a_0 = 2c^2\sqrt{\Lambda/3}$. This indeed leads to $(T_U - T_{\Lambda}) \to a$ when $a \gg a_0$ and $(T_U - T_{\Lambda}) \to a^2/a_0$ when $a \ll a_0$, and is the only example of an explicit theoretical/heuristic derivation of an interpolating function $\mu(x)=(\sqrt{1+4x^2}-1)/2x$. Similar ideas relying on dynamics in de Sitter spacetime have led to the proposals that MOND could arise either within a theory where spacetime and gravity both emerge together from the entanglement structure of an underlying microscopic theory \citep{Verlinde}, or as a novel regime of quantum gravity phenomena at temperatures below the de Sitter temperature \citep{Smolin}. Within a quantum description of a toy model for the Universe, containing only the cosmological constant and a localised matter source, and described by the coherent state of a massless scalar field, it has also been shown that the reaction of the de Sitter background to the presence of the matter source induces a MOND behaviour \citep{Giusti}. Hence if the MOND phenomenology in galaxies ends up teaching us some radically new physical principles, it is conceivable that these would be related to a deeper understanding of quantum phenomena in de Sitter spacetime.

\begin{ack}[Acknowledgments]

We thank Luc Blanchet, Harry Desmond, Jonathan Freundlich, Aurélien Hees, Federico Lelli, Mordehai Milgrom, and Tobias Mistele for reading a draft of this chapter and providing insightful comments. We thank Tobias Mistele for providing Fig.~3 and Constantinos Skordis for allowing us to present Fig.~5. B.F. acknowledges funding from the European Research Council (ERC) under the European Unions Horizon 2020 research and innovation program (grant agreement No. 834148) and from the Agence Nationale de la Recherche (ANR projects ANR-18-CE31-0006 and ANR-19-CE31-0017). A.D. was supported by the European Regional Development Fund and the Czech Ministry of Education, Youth and Sports: Project MSCA Fellowship CZ FZU I - CZ.02.01.01/00/22\_010/0002906. 
\end{ack}

\seealso{Review articles and books: \citet{Bruneton}, \citet{Famaey_2012}, \citet{Milgrom_2014}, \citet{Merritt}, \citet{Banik_2022}.}

\vspace{-0.4cm}

\bibliographystyle{Harvard}
\bibliography{reference}

\end{document}